\documentclass[prb,twocolumn,showpacs,preprintnumbers,superscriptaddress,amsmath,amssymb,10pt]{revtex4-2}

\usepackage[pdftex]{graphicx}
\usepackage{bm}

\usepackage{mathcomp}
\usepackage{amsmath}
\usepackage{placeins}

\usepackage{capt-of}

\usepackage{gensymb}

\usepackage{multirow}

\usepackage[justification=raggedright,singlelinecheck=false]{caption}

\begin{document}

\title{Semimetallic two-dimensional defective graphene networks with periodic 4-8 defect lines}
\author{Roland Gillen}\email{roland.gillen@swansea.ac.uk}
\affiliation{College of Engineering, Swansea University, Swansea SA1 8EN, United Kingdom }
\affiliation{Department of Physics, Friedrich-Alexander-Universit\"{a}t Erlangen-N\"{u}rnberg, Staudtstr. 7, 91058 Erlangen, Germany}
\author{Janina Maultzsch}
\affiliation{Department of Physics, Friedrich-Alexander-Universit\"{a}t Erlangen-N\"{u}rnberg, Staudtstr. 7, 91058 Erlangen, Germany}

\date{\today}

\begin{abstract}
We present theoretical simulations of the electronic properties of graphene-like two-dimensional (2D) carbon networks with a periodic arrangement of defect lines formed by alternating four- and eight-membered rings. 
These networks can be seen as arrays of armchair-edged nanoribbons (AGNRs), which are covalently connected at the edges. Using a combination of density functional theory and a simple tight binding model, we show that the electronic properties of these networks can be understood to arise from the family behaviour of the constituting AGNRs, plus a rigid shift due to an 'inter-ribbon' coupling across the defect lines. As a result, we find one class of zero-band-gap semiconducting 2D networks, and two classes of semimetallic networks with quasi-linear band close to the Fermi energy. The formation of closed-ring electron- and hole-like Fermi surfaces due to hybridization across the defect lines offers interesting perspectives of using such defective 2D networks for transport applications or the realization of carbon-based nodal line semimetals. 
\end{abstract}

\maketitle

\section{Introduction}
In recent years, the versatile chemical bonding capabilities of carbon has driven the exploration of carbon-based one- (1D) or two-dimensional (2D) lattices that exhibit metallic, semiconducting, or unconventional electronic characteristics, paving the way for the development of "all-carbon" structures. A cornerstone material in this context is graphene, which has demonstrated remarkable electrical conductivity, outstanding mechanical strength, and excellent thermal conductivity, making it a promising candidate for applications such as high-speed transistors, flexible displays, and energy storage devices~\cite{Schwierz2010,Novoselov2012,OLABI2021110026}. 
Quantum confinement effects and the introduction of edges or pores allow for a modification of the material characteristics, while retaining many of the promising electronic properties of graphene.

The introduction of structural defects, such as four- and eight-membered rings, provides an additional design dimension for carbon-based nanoribbons~\cite{Liu2017,Kang2023} and nanotubes~\cite{LI2018656}, as well as for novel two- and three-dimensional materials. In the latter class, recently proposed materials include T-carbon~\cite{PhysRevLett.106.155703}, T-graphene~\cite{QinyanGu.97401,T-graphene-nanotubes} and PHOTH-graphene~\cite{Santos2024}. 
In a previous work~\cite{Gillen_2024}, we studied the electronic properties of graphene nanoribbons with bisecting 4-8 ring defect lines. We found that, akin to defect-free armchair-edged nanoribbons~\cite{PhysRevLett.97.216803,PhysRevLett.130.026401} and carbon nanotubes, the defective ribbons can be classified into two families of semiconducting nanoribbons with bandgaps increasing due to quantum confinement for decreasing ribbon width, and a third family of nanoribbons featuring linear crossing bands at the Fermi energy. Using tight-binding models, we concluded that the electronic properties of the defective nanoribbons are defined by the properties of nanoribbon halves on either side of the defect lines, modified by an additional 'interhalf' coupling. Such defective nanoribbons could conceivably be realized through bottom-up synthesis techniques using suitable molecular precursors.  

In this paper, we extend our previous results~\cite{Gillen_2024} on defective nanoribbons to the case of graphene-like 2D networks with a periodic array of defect lines of alternating four- and eight-membered rings. We show that these 2D networks as well inherit the family properties of pristine AGNRs and can be categorized into three groups, according to their electronic bandstructures. The stronger hybridization effects across the defect lines lead to semi-metallic properties for two thirds of the networks, which might be interesting for transport applications and the realization of carbon-based nodal line semimetals.

\section{Computational Method}
The atomic geometries and electron bandstructures were calculated with the ABACUS computational package~\cite{abacus} using numerical atomic orbitals in combination with normconserving pseudopotentials from the SG15 library~\cite{sg15}. Energy cutoffs of 100\,Ry and 500\,Ry were used for the real-space representation of the electron wavefunctions and for the calculations of numerical orbital two-center integrals, respectively. We fully optimized the atomic positions as well as the lattice constant along the nanoribbon axis until the residual forces and stresses were below 0.0025\,eV/\AA\space and 0.01\,GPa, respectively. The atoms were allowed to move freely in both in-plane and out-of-plane directions and no symmetry constraints were imposed during the optimization step. For this step, the exchange-correlation interaction was approximated by the revised Perdew-Burke-Ernzerhof functional (PBEsol)~\cite{pbesol} and the electrons were represented by a triple-zeta double-polarized (TZDP) basis set. The two-dimensional Brillouin zone was sampled with $\Gamma$-centered k-points grid with a spacing of at least 0.1\,\AA$^{-1}$ (translating to a 16x16x1 k-point grid in case of a 4-d$_{48}$CN. We added vacuum layers of at least 30\,\AA\space thickness to minimize residual interactions between periodic images due to periodic boundary conditions. Using these parameters, the computed total energies are converged to within 1\,meV/atom (cf. supplemental information). To ensure accurate predictions of the electronic properties, we then calculated the electronic band dispersions using the HSE12 hybrid functional~\cite{hse12} and a DZP basis set for the PBEsol-level atomic structure.

\section{Results and Discussion}
\begin{figure}
	\centering
	\includegraphics*[width=0.99\columnwidth]{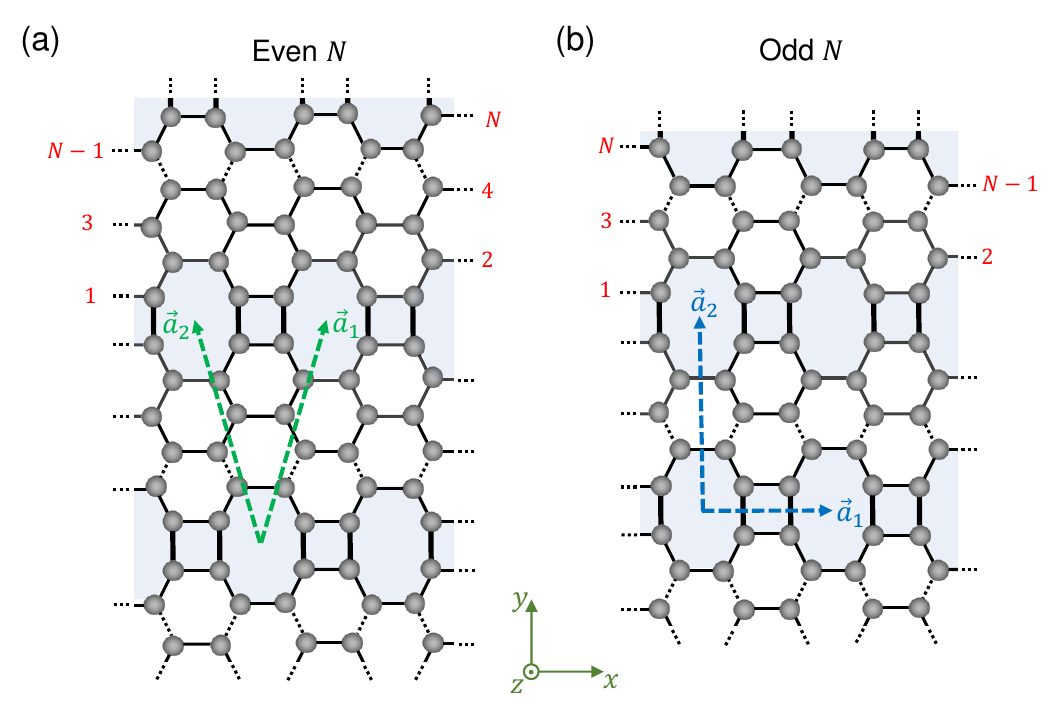}
	\caption{\label{fig:structure_dAGNR} (Color online) Atomic structures of graphenic networks with 4-8-ring defect lines. (a) Structure for 2D networks with \textit{even} number $N$ of carbon dimer lines between two defect lines. (b) Structure of 2D networks with \textit{odd} $N$. The lattice vectors $a_1$ and $a_2$ of the 2D networks are shown as dashed arrows. The bonds across the defect lines are elongated and indicated through thicker black lines.}
\end{figure}
\begin{figure*}
	\centering
	\includegraphics*[width=0.49\textwidth]{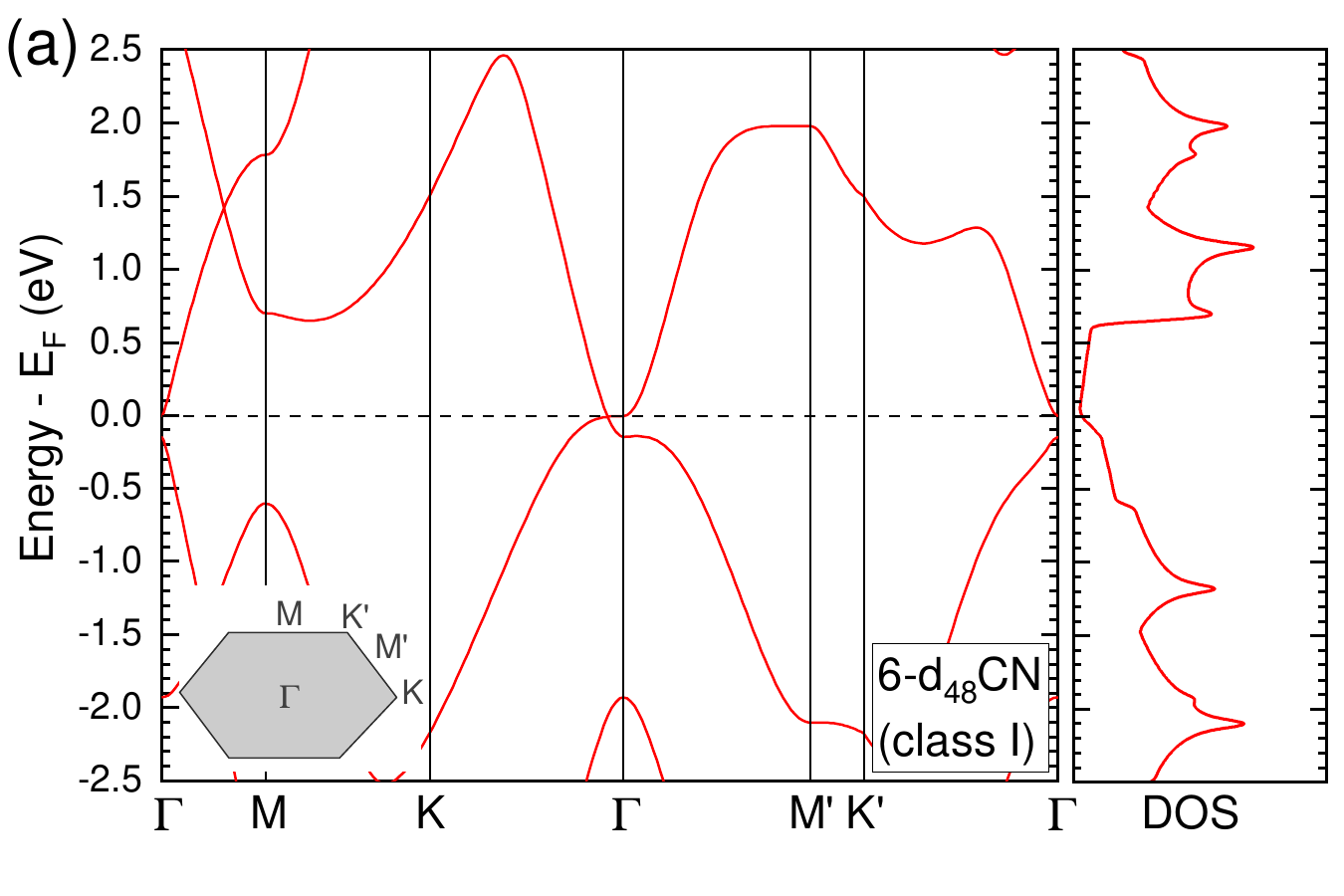}
	\includegraphics*[width=0.49\textwidth]{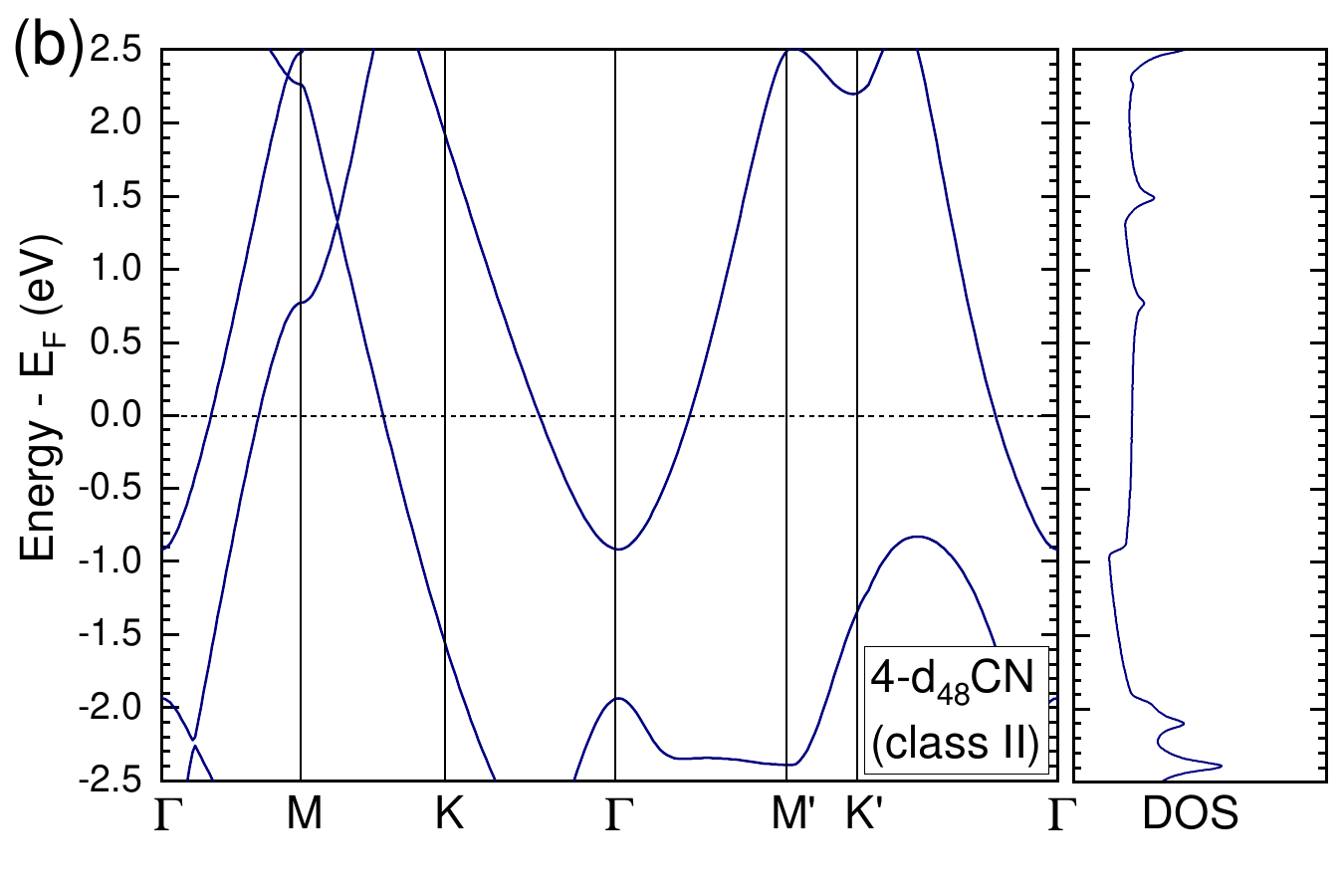}
	\includegraphics*[width=0.49\textwidth]{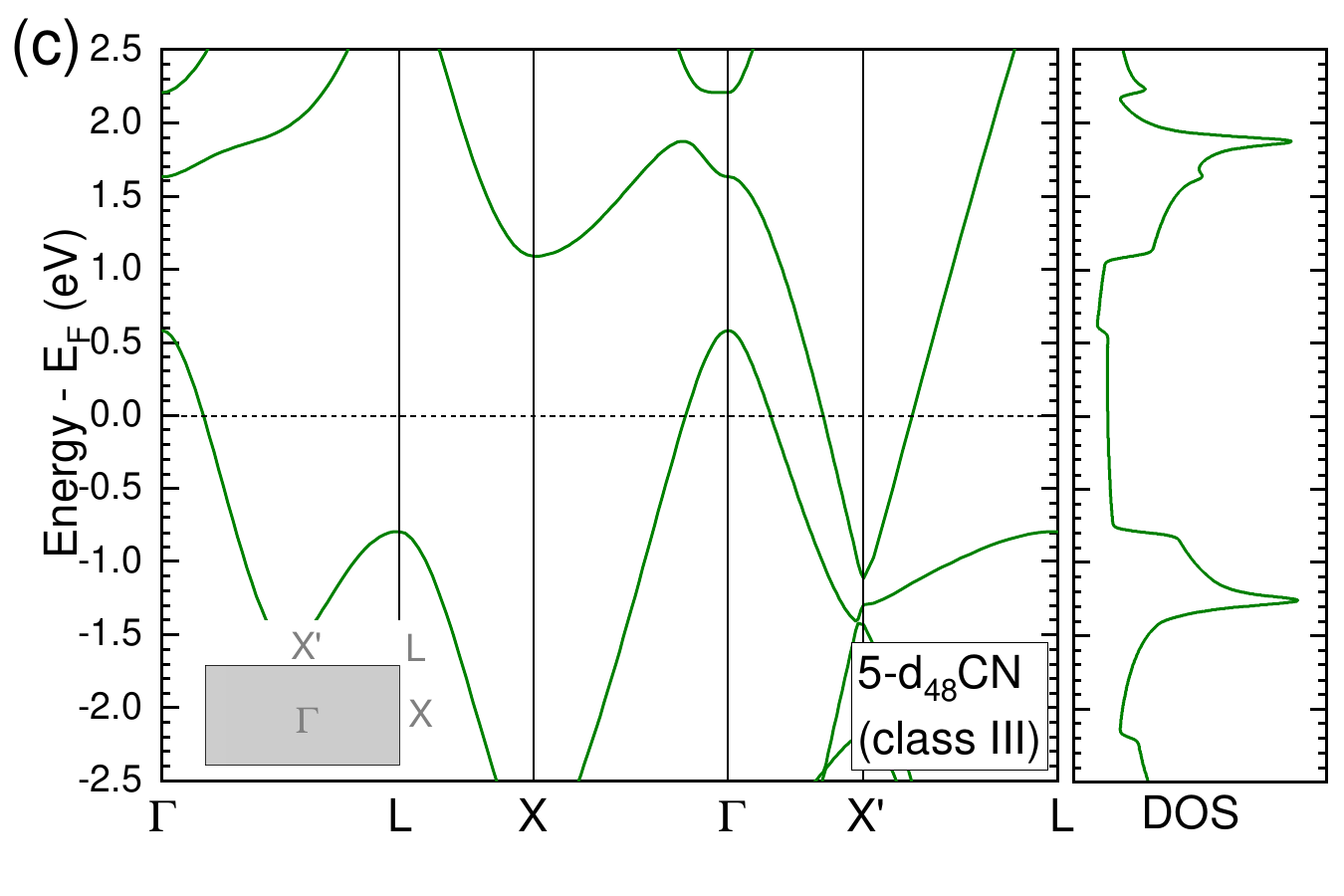}
	\includegraphics*[width=0.49\textwidth]{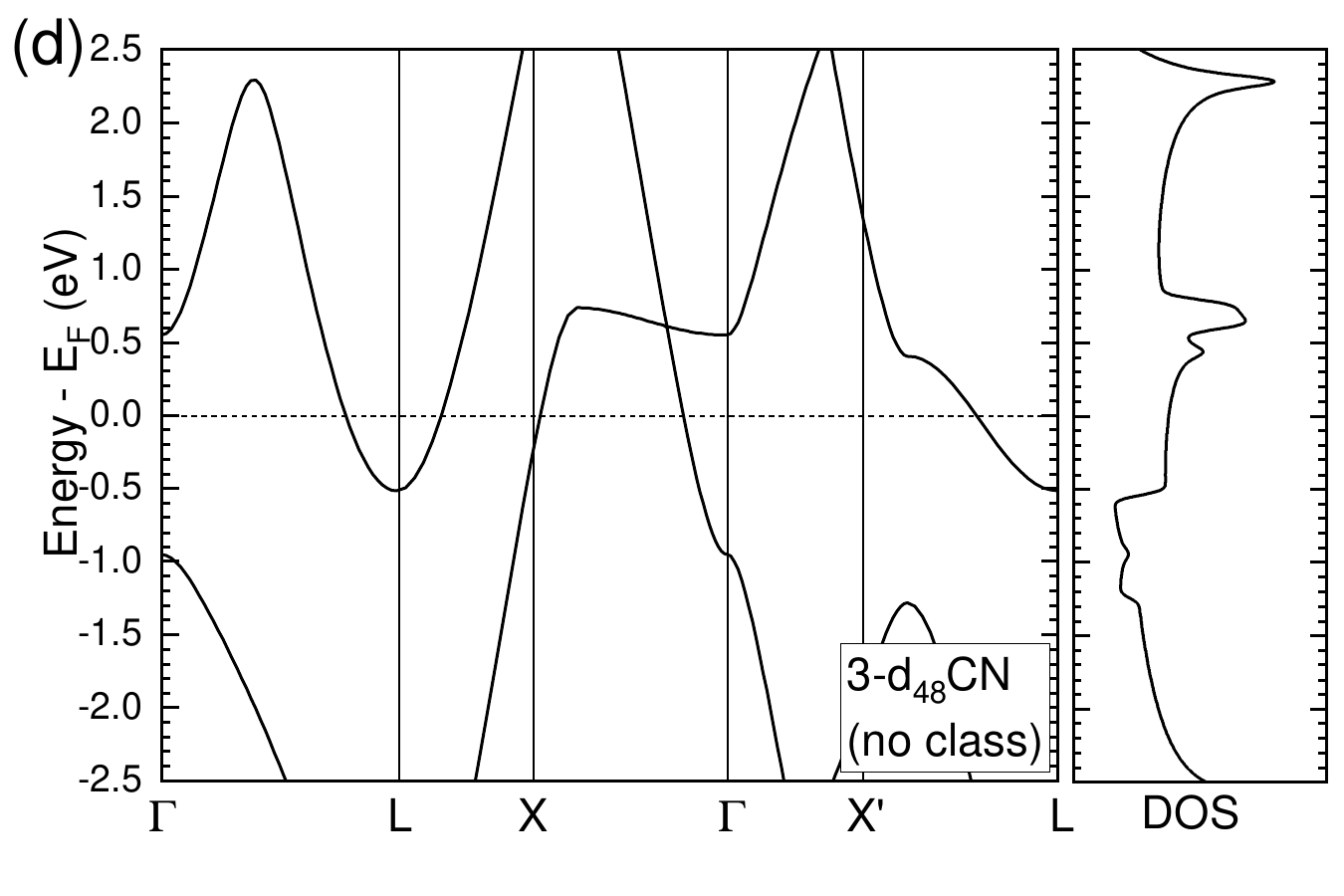}
	\caption{\label{fig:bands_dCN} (Color online) (a)-(c) Computed electronic bandstructures of $N$-d$_{48}$CN for $N$=4-6 as representative members of the three defective network groups. (d) Electronic bandstructure of the recently synthesized 3-d$_{48}$CN, which does not fit into the class behavior. Note that for networks, even (odd) $N$, the primitive cells generally are rhombohedral (rectangular). The grey dashed lines indicate the Fermi energy of the systems. Insets schematically show the Brillouin zones of networks with even and odd $N$. All results were obtained on the HSE12-level of theory.}\end{figure*}
\begin{figure}
	\centering
  \includegraphics*[width=0.49\columnwidth]{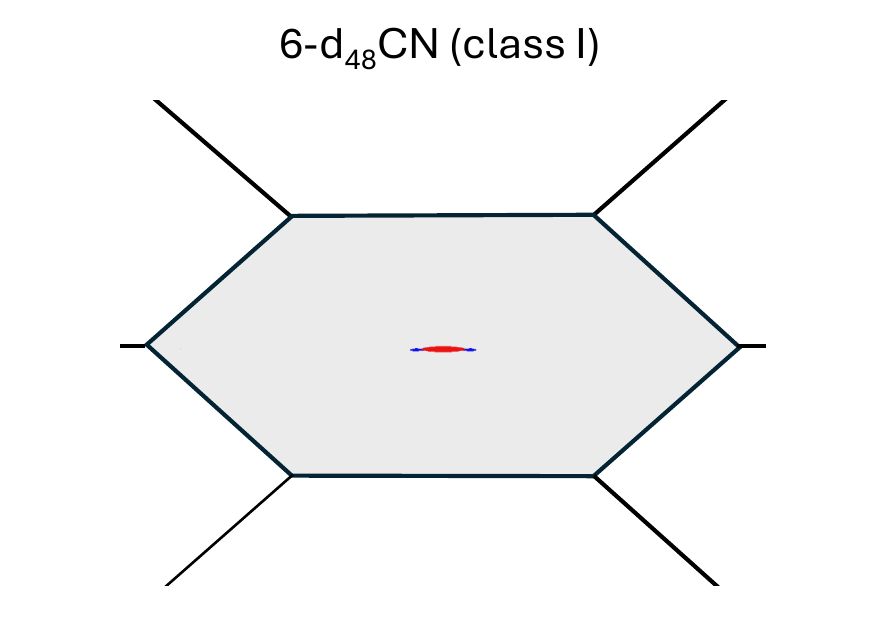}
	\includegraphics*[width=0.49\columnwidth]{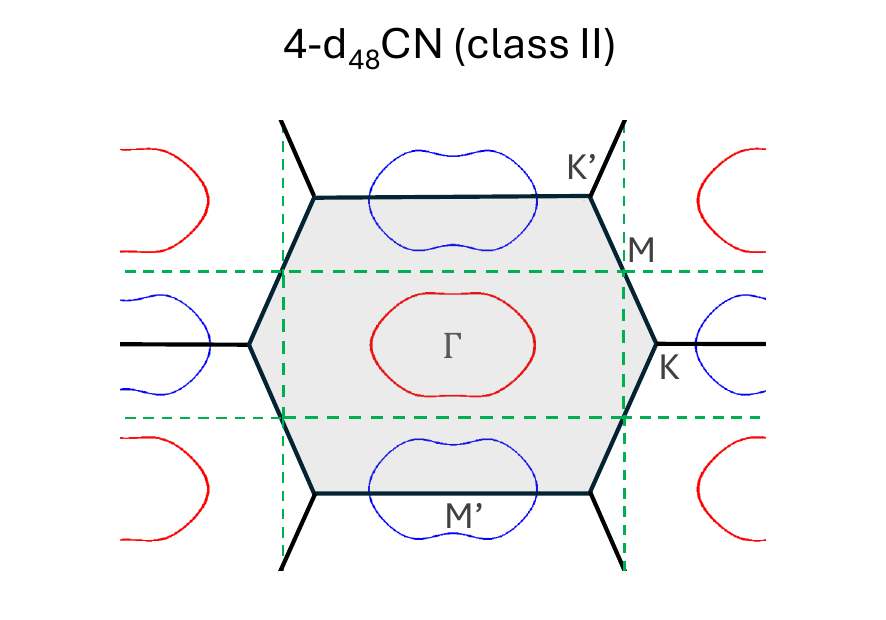}
	\includegraphics*[width=0.49\columnwidth]{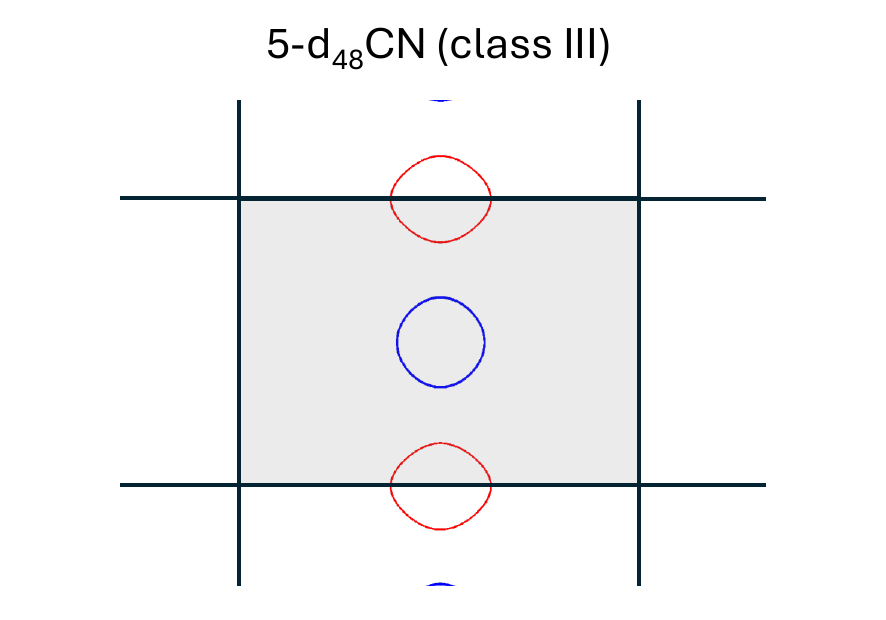}
	\includegraphics*[width=0.49\columnwidth]{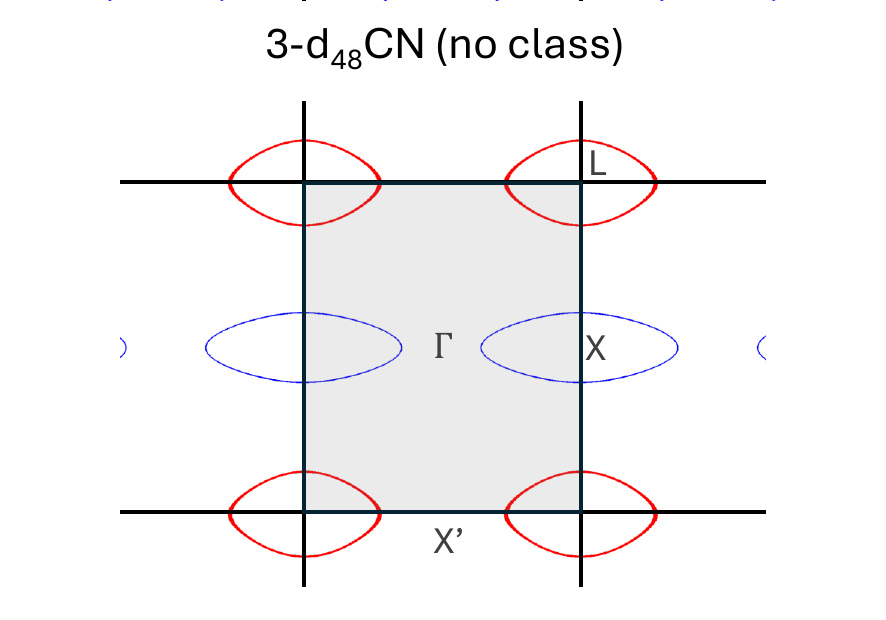}
	\caption{\label{fig:fermi_surface} (Color online) Fermi surfaces of a $N$-d$_{48}$CN with $N$=3-6 [cf. Fig.~\ref{fig:bands_dCN}~(a)-(d)]. Red and blue  color represent electron and hole contributions to the Fermi surface, respectively. The Brillouin zone corresponding to the primitive cell is shown as shaded grey area. The Brillouin zones corresponding to the \textit{conventional} cell (dashed green lines) are shown for the $N$-d$_{48}$CN.}
\end{figure}
Based on our previous study~\cite{Gillen_2024} of 1D graphene nanoribbons with a single 4-8-defect line, we here consider two-dimensional graphenic sheets with periodic 4-8-ring defect lines. These can be seen as arrays of covalently connected nanoribbons. We classify the networks according to the number $N$ of carbon dimer lines along the axis of each constituting nanoribbon. For odd $N$, the networks adopt a centered rectangular lattice (space group $D_{2h}^{19}$), for even $N$ the networks adopt a primitive rectangular structure (space group $D_{2h}^1$), cf. Fig.~\ref{fig:structure_dAGNR}. In a first step, we used density functional theory simulations (refer to Sec.~\ref{sec:method} for computational details) to fully optimize the atomic structures for defective 2D networks with $N=3-18$. As we found previously~\cite{Gillen_2024} for defective nanoribbons, the defect lines distort the bond lengths close to the defect lines; the bond lengths next to the defect line are shortened, similar to the bond lengths at the edges of armchair graphene nanoribbons. For sufficient width of the nanoribbons between the defect lines, the bond lengths approach the computed value in defect-free graphene (1.423\,\AA). The bond lengths across the defect lines are significantly elongated, with values of around 1.5\,\AA. Despite the presence of the defect lines composed of non-benzenoid rings, our simulation predict all networks to be flat. We further calculated the phonon dispersion of the smallest networks and found no imaginary modes, suggesting that all networks are dynamically stable [cf. supplemental material].

Based on the optimized structures, we simulated the electronic bandstructures for the defective networks. Figure~\ref{fig:bands_dCN}~(a)-(c) shows the bandstructures for $N=4-6$, which are representative for the entire set of considered defective networks. The calculated bandstructures for networks with $N=3-17$ on the PBEsol-level of theory are shown in Fig.~S3 of the supplemental material.
We can distinguish three different classes of networks with 4-8-ring defect lines: Class I networks are composed of $M$-AGNRs of the $M=3p$ family (with integer $p$), which possess relatively wide band gaps~\cite{PhysRevLett.97.216803}. In this case, refer to the representative bandstructure of a 6-d$_{48}$CN shown in Fig.~\ref{fig:bands_dCN}~(a), we observe a crossing of valence and conduction band at a point on the $\Gamma-K$ (even $N$) or $\Gamma-X$ lines (odd $N$), with a flat band feature at the Fermi energy between this crossing point and $\Gamma$. As we will discuss in more detail later, the formation of this band feature can be understood from first-nearest neighbor coupling across the defect line. The crossing point of the valence and conduction band coincides with the Fermi energy and the Fermi surface is reduced to two single points. The class I networks hence are predicted to be zero-band-gap semiconductors.

For class II networks [Fig.~\ref{fig:bands_dCN}~(b)], the $\Gamma$ point conduction band minimum is below the Fermi energy, forming an electron pocket, while the Fermi energy crosses the valence band in the vicinity of $X'$ (odd $N$) or $M$ (even $N$). Both valence and conduction bands are approximately linear close to the Fermi energy. The situation is reversed for class III networks, see Fig.~\ref{fig:bands_dCN}~(c). The C-C bond lengths across the defect lines show a weak class structure with periodicity $T_N=3$ as well, cf. Fig.~S2 in the supplemental material. 

Figure~\ref{fig:bands_dCN}~(d) shows the bandstructure of a 3-d$_{48}$CN, which has been recently realized through bottom-up synthesis from bi-phenyl precursors~\cite{Fan-2024} and was found to be ultraflat and metallic, in agreement with our DFT-based predictions. This network does not appear to fit into the periodic class trends of the networks $N\leq 4$, with the conduction band crossing the Fermi energy around the $L$ point and a hole pocket around the $X$ point. The deviation from the class behavior might be caused by a larger contribution of the defect lines over the thin hexagonal ring segments.

As for the defective 1D nanoribbons~\cite{Gillen_2024}, these electronic properties can be understood from the reduced coupling of the smaller $N$-AGNRs constituting the network across the defect lines. The coupling gives rise to a hybridization of the individual nanoribbon bands, which in case of the defective nanoribbons resulted in a reduction of the electronic band gap and, for two of the three nanoribbon families, to a direct-to-indirect band gap transition with either the valence or the conduction band extremum moving away from the $\Gamma$ point. The stronger hybridization in the 2D networks causes a semiconductor-to-metal transition, while retaining a one-to-one correspondence between the three network classes and the three defect-free AGNR families.

We corroborated these qualitative arguments with explicit tight-binding simulations, where we treated the network as consisting of individual 'nanoribbons', coupled by a Hamiltonian 
\begin{equation}
	H_{cpl} = \gamma\sum_{\left<i,j\right>}\left(\hat{a}_{i}^\dagger\hat{a}_{j}\right)\label{eq:Hint}
\end{equation}
$\gamma$ couples nearest-neighbor sites $i$ and $j$ across the defect lines (i.e. $i$ and $j$ are the atoms making of the four-membered rings) and is zero otherwise. 
For the electronic structure of the individual nanoribbons making up the networks, we included interactions up to three-nearest neighbors. We found that this simple model yields electronic bandstructures that are in good qualitative agreement with the DFT results~\footnote{The inclusion of interactions across the defect line up to three nearest-neighbors improves the quantitative agreement with the DFT results but requires a more elaborate choice of the hopping parameters.} We then varied $\gamma$ across the defect lines to study its effect on the network electronic properties. 

For vanishing coupling strength, the bandstructure of each network is a pure superposition of the bandstructures of the individual AGNRs in the unit cell. In case of the class I networks, the direct band gap at the $\Gamma$ point is reduced due to hybridization and the valence band $v+1$ approaches the valence band top as the 'inter-ribbon' coupling $\gamma$ is gradually increased. For sufficiently large coupling strength, the conduction band $c$ and valence band $v+1$ cross the Fermi energy. This gives rise to a flat band dispersion close to the Brillouin zone center in $\Gamma-X$ (odd $N$) or $\Gamma-K$ direction, but not in the other high symmetry directions.

For class II networks, which consist of AGNRs of the family $N=3p+1$, our simulations suggest that the 'inter-ribbon' coupling first causes a transformation of the band gap from direct to indirect, as the conduction band minimum at $\Gamma$ and the valence band at $X'/M$ shift to lower and higher energies, respectively. For sufficiently large coupling strength, a semimetallic state with an electron pocket at $\Gamma$ and a hole pocket at $X'/M$ arises, as indicated in Fig.~\ref{fig:bands_dCN}~(a) for a 4-d$_{48}$CN. 
For class III networks, we observe the opposite behavior: these networks consist of AGNRs of the family $N=3p+2$, which, in absence of edge effects, feature a vanishing band gap at $\Gamma$ in tight-binding simulations. Here, we observe that the inter-ribbon coupling causes the valence band to cross the Fermi level around the $\Gamma$ point, while an electron pocket forms at the $X'/M$ point. For isolated AGNRs of this family, the vanishing band gap can be understood from zone-folding arguments: the $K$ point of graphene is mapped onto the $\Gamma$ point of the nanoribbon; consequently, the nanoribbon bandstructure close to the Fermi energy corresponds to a cut through one of the graphene Dirac cones, resulting in the characteristic linear bands. As the bandstructure of the defective networks close to the Fermi energy is determined by the electronic structure of the underlying AGNRs plus a hybridization-induced rigid shift, this suggests a large mobility of the electrons close to the Fermi energy. 
\begin{figure}
	\centering
	\includegraphics*[width=0.95\columnwidth]{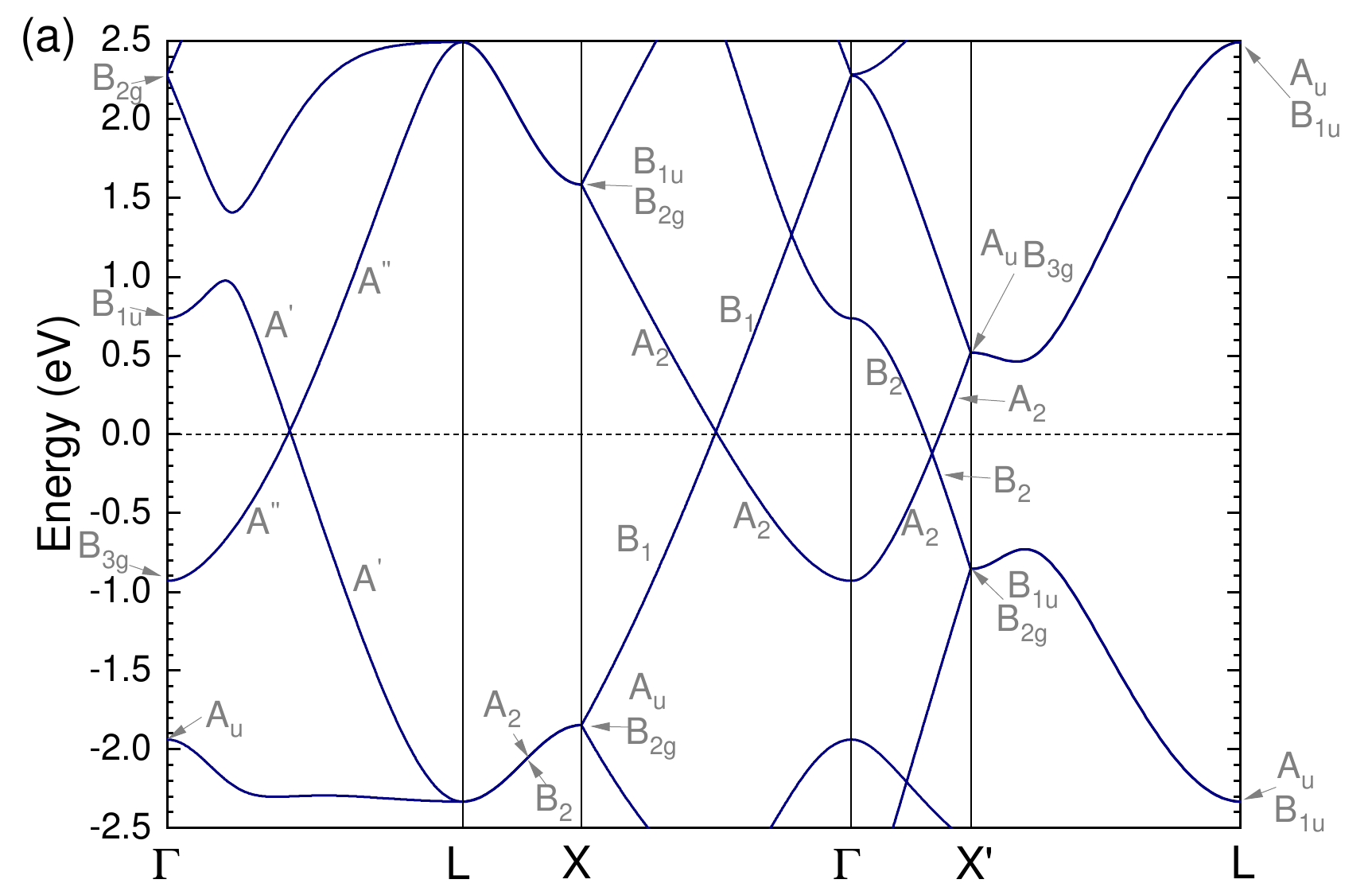}
	\includegraphics*[width=0.95\columnwidth]{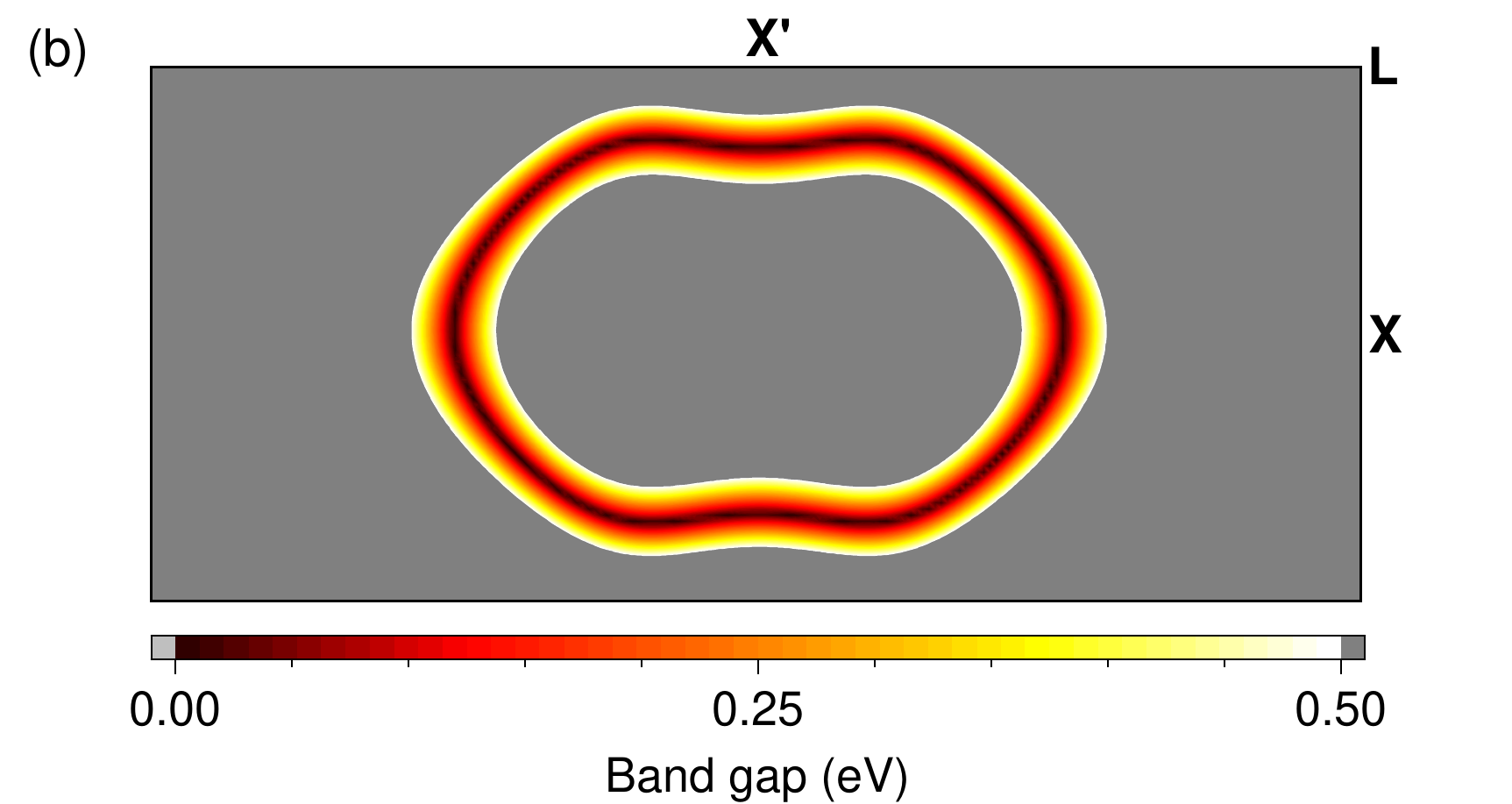}
	\caption{\label{fig:4dCN_conv_gaps} (Color online) (a) Electronic bandstructure of a 4-d$_{48}$CN for a small z-offset between neighboring nanoribbons. The unit cell is the conventional cell of the undistorted network and contains two defect lines. Several quasi-linear crossing bands appear near the Fermi energy for paths between the $\Gamma$ point and the high-symmetry points. The symmetries of the relevant bands around the Fermi energy are indicated. (b) Band gap between the crossing valence and conduction bands throughout the two-dimensional Brillouin zone, revealing a Dirac nodal line. All results from simulations with the HSE12 functional.}
\end{figure}
\begin{figure}
	\centering
	\includegraphics*[width=0.9\columnwidth]{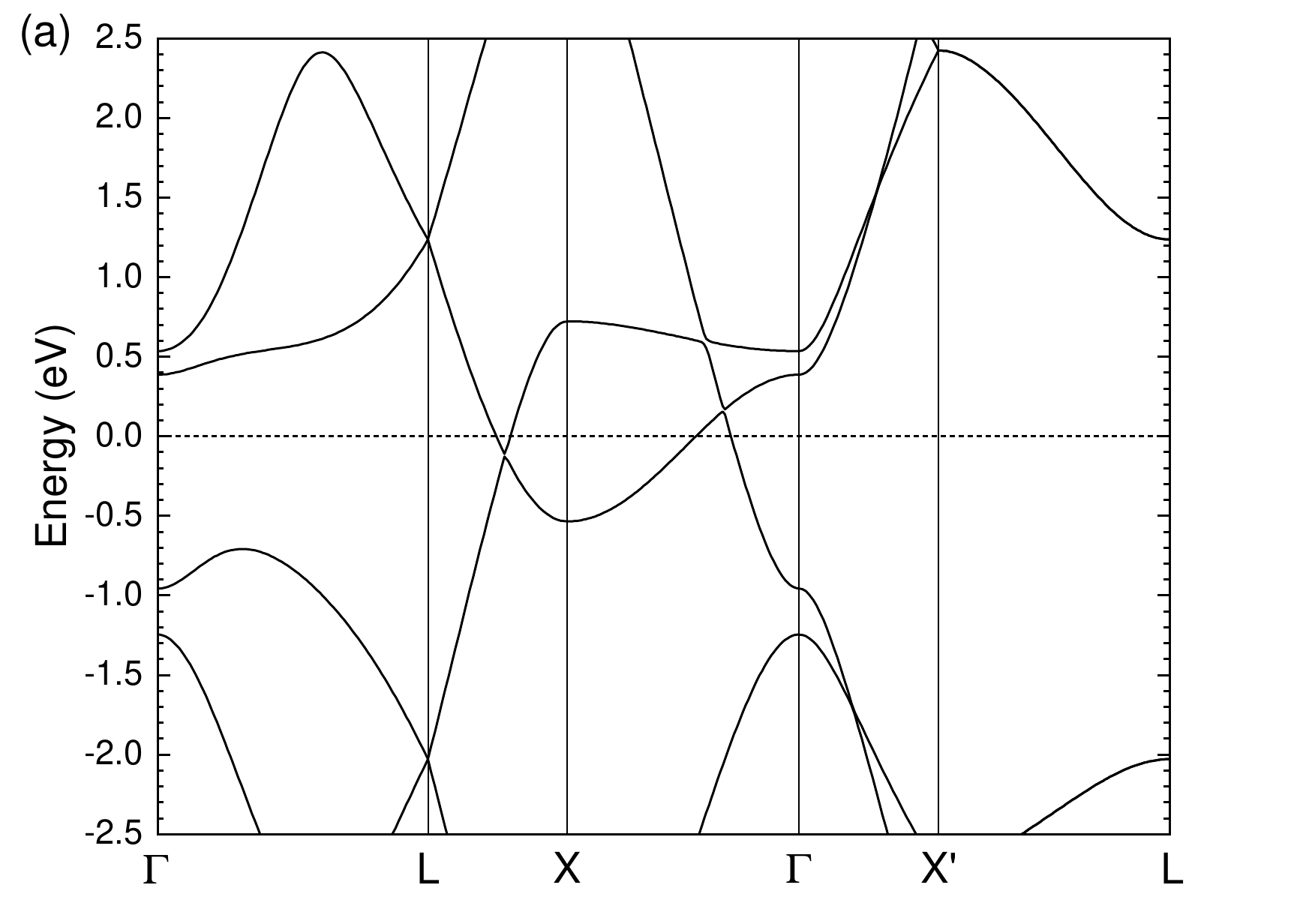}
	\includegraphics*[width=0.9\columnwidth]{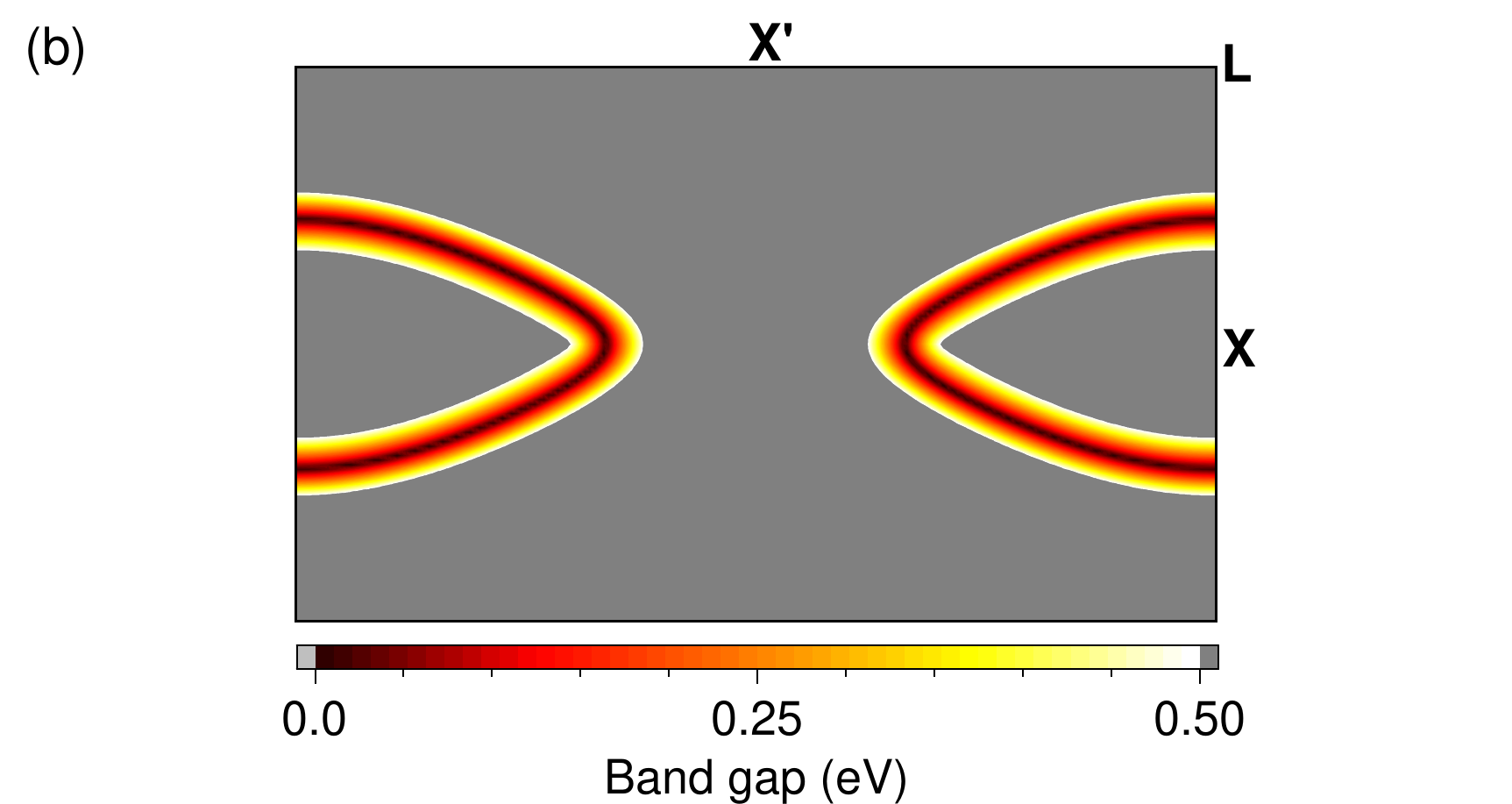}
	\caption{\label{fig:3dCN_conv_gaps} (Color online) (a) Electronic bandstructure of a 3-d$_{48}$CN with a small z-offset between neighboring nanoribbons. The unit cell is thus a 1x2x1 of the rectangular primitive cell of the unperturbed network. (b) Band gap between the crossing valence and conduction bands throughout the two-dimensional Brillouin zone, with a Dirac nodal line centered around the $X$ point. All results from simulations with the HSE12 exchange-correlation functional. }
\end{figure}

Interestingly, for networks of class II, the Fermi energy crosses the conduction band along each of the three high-symmetry paths involving the $\Gamma$ point, while the same is true for the valence band in case of class III networks. This suggests that the Fermi surface forms a ring around the $\Gamma$ point. On the other hand, crossings of the valence (class II) or conduction bands (class III) occur mainly around the M (even $N$) or X' (odd $N$) points. To further analyse this, we computed the Fermi surface for representative networks. Fig.~\ref{fig:fermi_surface} shows the Fermi surface for a 4-d$_{48}$CN. As expected, the conduction band gives rise to a closed ring centered around the $\Gamma$ point in the Fermi surface. The valence band, on the other hand, forms a corresponding closed ring around the $M$ point, but not $M'$. We find similar results for all class II and class III networks, and also the 'classless' 3-d$_{48}$CN. 

The origins and locations of the two rings in the Fermi surface can be understood from a simple tight-binding argument:
The wavefunction of band $\lambda$ at k-point $\vec{k}$ in terms of the tight-binding basis $\phi_{m,\vec{T}}\left(r\right)$ is given by
\begin{equation}
	\psi_{\lambda k}(\vec{r}) = \sum_{\vec{T}}c_{\lambda,m}\left(\vec{k},\vec{T}\right)e^{i\vec{k}\cdot\vec{T}}\phi_{m}\left(\vec{r}-\vec{T}\right)\label{eq:eq1}
\end{equation}
where $\vec{T}$ is a lattice vector. In the limit of vanishing 'inter-ribbon' coupling $\gamma$, the coefficients c$_{\lambda,m}$ only depend on the components of $\vec{T}$ perpendicular to the defect lines. This results in a completely flat band dispersion for all Brillouin zone paths that are perpendicular to the defect lines, e.g. the $\Gamma-X'$ path for odd $N$. 

In first-order perturbation theory, the energy correction from $H_{cpl}$ defined in Eq.~\ref{eq:Hint} to band $\lambda k$ with wavefunction $\psi_{\lambda k}^0$ of the network without inter-ribbon coupling is
\begin{equation}
	\Delta\epsilon_\lambda(\vec{k}) = \left<\psi_{\lambda k}^0|H_{cpl}|\psi_{\lambda k}^0\right>.
\end{equation}
For simplicity, we will first consider the case of odd $N$. $H_{cpl}$ can then be seen to couple nearest-neighbors across the defect line that are located in neighboring primitive cells in y ($\vec{a}_2$) direction. This introduces a phase factor due to the exponential term in Eq.~\ref{eq:eq1} For the energy corrections of the highest valence ($v$) and lowest conduction band ($c$), this implies
\begin{equation}
	\Delta\epsilon_{v/c}(X') = \exp\left(i\pi\right)\Delta\epsilon_{v/c}(\Gamma).
\end{equation}
Combined with the flat dispersion in $\Gamma-X'$ direction in case of $\gamma\rightarrow 0$, a filling of the conduction band around $\Gamma$ due to hybridization across the defect lines is compensated for by an equal partial depletion of the valence band around the $X'$ point. Similar arguments hold for the case of even $N$. In contrast to graphene, the two closed loops in the Fermi surface result in a significant density of states of ultramobile charge carriers at the Fermi level.

If we assume for a moment a small structural modulation in perpendicular direction to the defect lines (the $y$-direction in Fig.~\ref{fig:structure_dAGNR}) that removes the degeneracy between neighboring nanoribbons, the $M$ (even $N$) or $X'$ (odd $N$) point would be folded onto $\Gamma$, superimposing the two rings [cf. Fig.~\ref{fig:fermi_surface}].  To illustrate the effect of such a weak symmetry breaking, Fig.~\ref{fig:4dCN_conv_gaps}~(a) shows the bandstructure of a 4-d$_{48}$CN for a network with a small $z$-offset of 0.15\,\AA\space between neighboring nanoribbon segments. The unit cell of this perturbed network is the rectangular conventional cell containing two defect lines and the \textit{symmorphic} $\sigma_h$ inplane mirror plane is turned into a \textit{nonsymmorphic} symmetry operation with an additional required shift by $\frac{1}{2}\vec{a}_2$. Due to zonefolding and the aforementioned anti-symmetry in the inter-ribbon coupling-induced energy corrections at $\Gamma$ and $M$, the bandstructure exhibits quasi-linear crossing bands at the Fermi energy. Plotting the energy difference between the top valence and the lowest conduction band, Fig.~\ref{fig:4dCN_conv_gaps}~(b), suggests that the bands in fact form a quasi-Dirac nodal line with band gaps smaller than 50\,meV (on the PBEsol level of theory). We find similar results for all class II and III nanoribbons. For the 3-d$_{48}$CN, a similar perturbation leads to a nodal line centered around the $X$ point of the rectangular Brillouin zone [cf. Fig.~\ref{fig:3dCN_conv_gaps}].

Nodal-line semimetals (NLSM) have received attention recently due to their transport properties and the potential for topological boundary states~\cite{NLSM-review}. In these materials, nodal lines of crossing linear bands arise from a band inversion mechanism in combination with a topological protection of the crossings due to crystal symmetry. In case of our perturbed networks, the Dirac nodal lines are allowed by the \textit{non-symmorphic} $(\sigma_h|0,\frac{1}{2}\vec{a}_2,0)$ symmetry operation, which leads to differing symmetries of the crossing bands [cf. Fig.~\ref{fig:4dCN_conv_gaps}~(a).
However, most reported nodal line semimetals adopt a hexagonal lattice, while examples of 2D nodal line semimetals of square or rectangular symmetry are rare. Further, many previously proposed NLSM contain heavy atoms and the related spin-orbit coupling (SOC) induces relatively large bands gaps of order 100$\,$meV in the nodal lines, e.g. in a Cu$_2$Si monolayer~\cite{Feng2017}. For this reason, NLSM consisting of lighter elements and a weaker SOC are of particular interest. For instance, a carbon nitride covalent 2D network~\cite{C9N4-NLSM} and Al$_2$B$_2$ and AlB$_4$ monolayers~\cite{AlB-NLSM} have been recently proposed as low-band gap NLSMs with exotic electronic properties. For our networks as well, SOC effects change the band gaps by less than 1\,meV. Additionally, several purely carbon-based three-dimensional NLSMs have been proposed and studied based on density functional theory simulations~\cite{LI201839,Li2019,PhysRevB.107.245111}, but, to the best of our knowledge, experimental realizations of these materials still have not been achieved.

\section{Conclusion}
In summary, we used density functional theory to study graphene-like carbon networks with a regular arrangement of 4-8-ring defect lines. We showed that the band structures of these 2D networks are largely dominated by the family structure of the nanoribbon segments between the defect lines, which are modified by hybridization effects across the defect line. As a result, two out of three defective network classes are semimetallic, with quasi-linear bands crossing the Fermi energy. The significant density-of-states of electron and hole charge carriers, which retains much of the outstanding charge carrier mobility of graphene, renders these defective networks potentially interesting for transport applications. The impact of 'inter-ribbon' hybridization indicates a tunability of the network electronic properties through strain in perpendicular direction to the defect line. Synthesis of such networks would be feasible using bottom-up techniques through suitable molecular precursors.  
Further, modifying the periodicity of the presented defective networks in perpendicular direction to the defect lines might serve as a starting platform for the experimental realization of purely carbon-based nodal-line semimetals with rectangular symmetry and small band gaps. In this direction, further work on the detailed symmetry and topological properties of the electronic bandstructure is necessary to confirm the nature of the nodal line band crossings.

\section{Acknowledgements}
The authors gratefully acknowledge the scientific support and HPC resources provided by the Erlangen National High Performance Computing Center (NHR@FAU) of the Friedrich-Alexander-Universität Erlangen-Nürnberg (FAU) under the NHR project \texttt{b181dc}. NHR funding is provided by federal and Bavarian state authorities. NHR@FAU hardware is partially funded by the German Research Foundation (DFG) – 440719683. This work was supported by the Deutsche Forschungsgemeinschaft (DFG, German Research Foundation) – GRK2861 – 491865171. We further acknowledge the support of the Supercomputing Wales project, which is part-funded by the European Regional Development Fund (ERDF) via the Welsh Government.


%


\pagebreak
\clearpage
\widetext
\renewcommand{\floatpagefraction}{0.99}
\begin{center}
\textbf{\Large Supplementary Material}\\
\textbf{\Large Semimetallic two-dimensional defective graphene networks with periodic 4-8 defect lines}\\
\vspace{0.8cm}
{\large Roland Gillen$^{1,2}$ and Janina Maultzsch$^{2}$}\\
\vspace{0.2cm}
{$^1$ College of Engineering, Swansea University, Swansea SA1 8EN, United Kingdom}
{$^2$ Department of Physics, Friedrich-Alexander University Erlangen-N\"{u}rnberg, Staudtstr. 7, 91058 Erlangen, Germany}
\end{center}

\setcounter{equation}{0}
\setcounter{figure}{0}
\setcounter{table}{0}
\setcounter{page}{1}
\setcounter{section}{0}
\makeatletter
\renewcommand{\theequation}{S\arabic{equation}}
\renewcommand{\thefigure}{S\arabic{figure}}
\renewcommand{\bibnumfmt}[1]{[S#1]}
\renewcommand{\citenumfont}[1]{S#1}
\renewcommand{\floatpagefraction}{.98}%

\section{Lattice constants of $N$-d$_{48}$CNs}
The table shows the evolution of the lattice constants of the defective networks as the distance between defect lines is increased (i.e. increase of $N$). All values are given for the rectangular conventional cell, i.e. containing two primitive cells in case of even $N$. $a$ and $c$ are the lattice constants parallel and perpendicular to the defect lines. In case of even $N$, the primitive lattice vectors are given by
\begin{align}
\vec{a}_1 &= ( a/2, c/2, 0)\nonumber\\
\vec{a}_2 &= (-a/2, c/2, 0)\nonumber
\end{align}

\begin{center}
\begin{tabular}{c c c c}
\hline
\hline
 $N$  &  Class & lattice constant $a$  & lattice constant $c$\\
      &         & (in $\AA$)            & in ($\AA$)         \\
\hline
3	  &	None &	4.511 &	3.754	\\
4	  &	II	 &	4.426	&	10.024	\\
5	  &	III	 &	4.380	&	6.263	\\
6	  &	I	   &	4.354	&	15.016	\\
7	  &	II	 &	4.342	&	8.732	\\
8	  &	III	 &	4.331	&	19.938	\\
9	  &	I	   &	4.320	&	11.206	\\
10	&	II	 &	4.314	&	24.872	\\
11	&	III	 &	4.310	&	13.665	\\
12	&	I	   &	4.305	&	29.807	\\
13	&	II	 &	4.301	&	16.132	\\
14	&	III	 &	4.300	&	34.728	\\
15	&	I	   &	4.296	&	18.595	\\
16	&	II	 &	4.294	&	39.659	\\
17	&	III	 &	4.292	&	21.055	\\
18	&	I	   &	4.291	&	44.585	\\
\hline
\hline
\end{tabular}
\end{center}

\FloatBarrier

\section{Phonon dispersion of 4-d$_{48}$AGNR}
\begin{figure}
\centering
\includegraphics*[width=0.32\columnwidth]{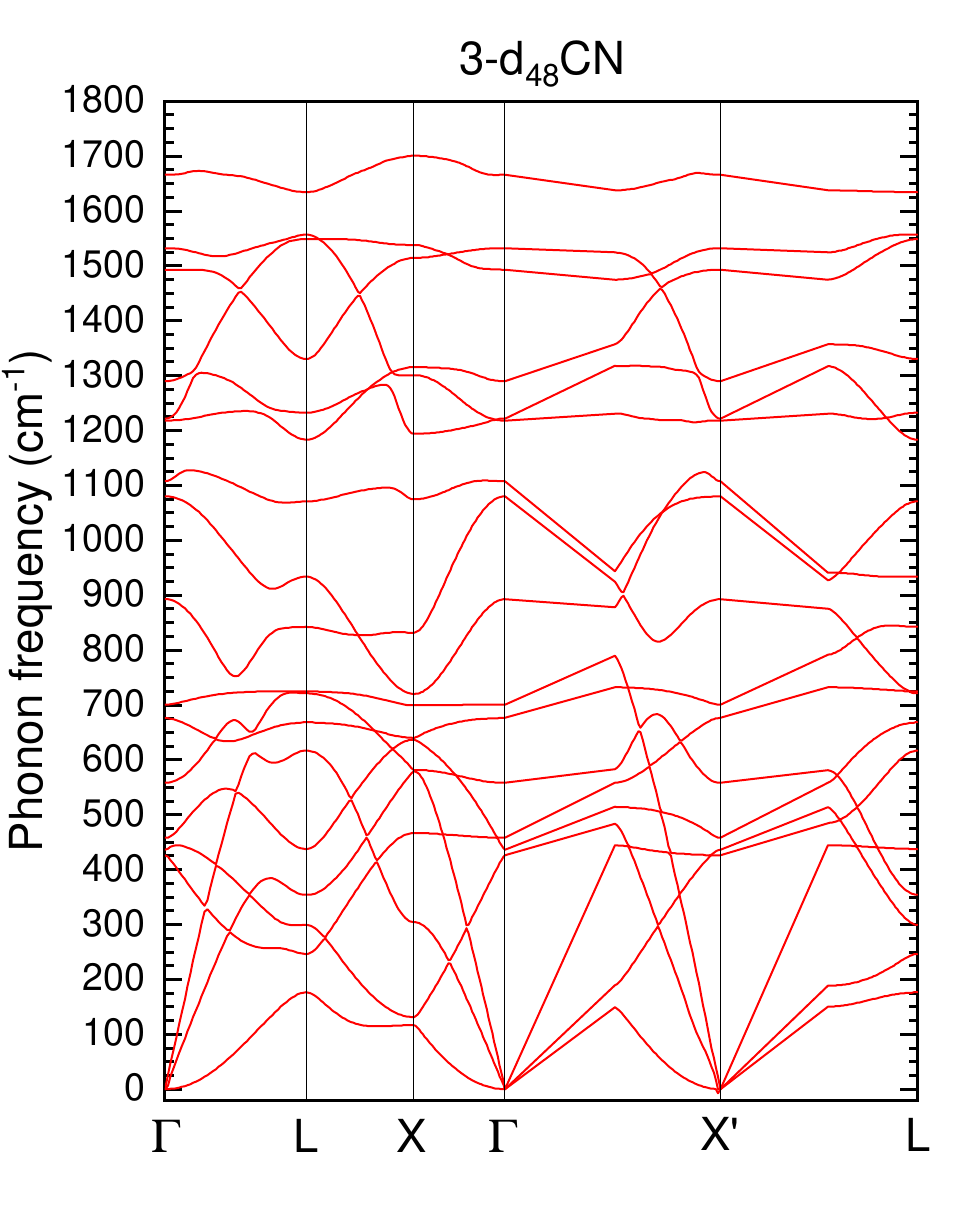}
\includegraphics*[width=0.32\columnwidth]{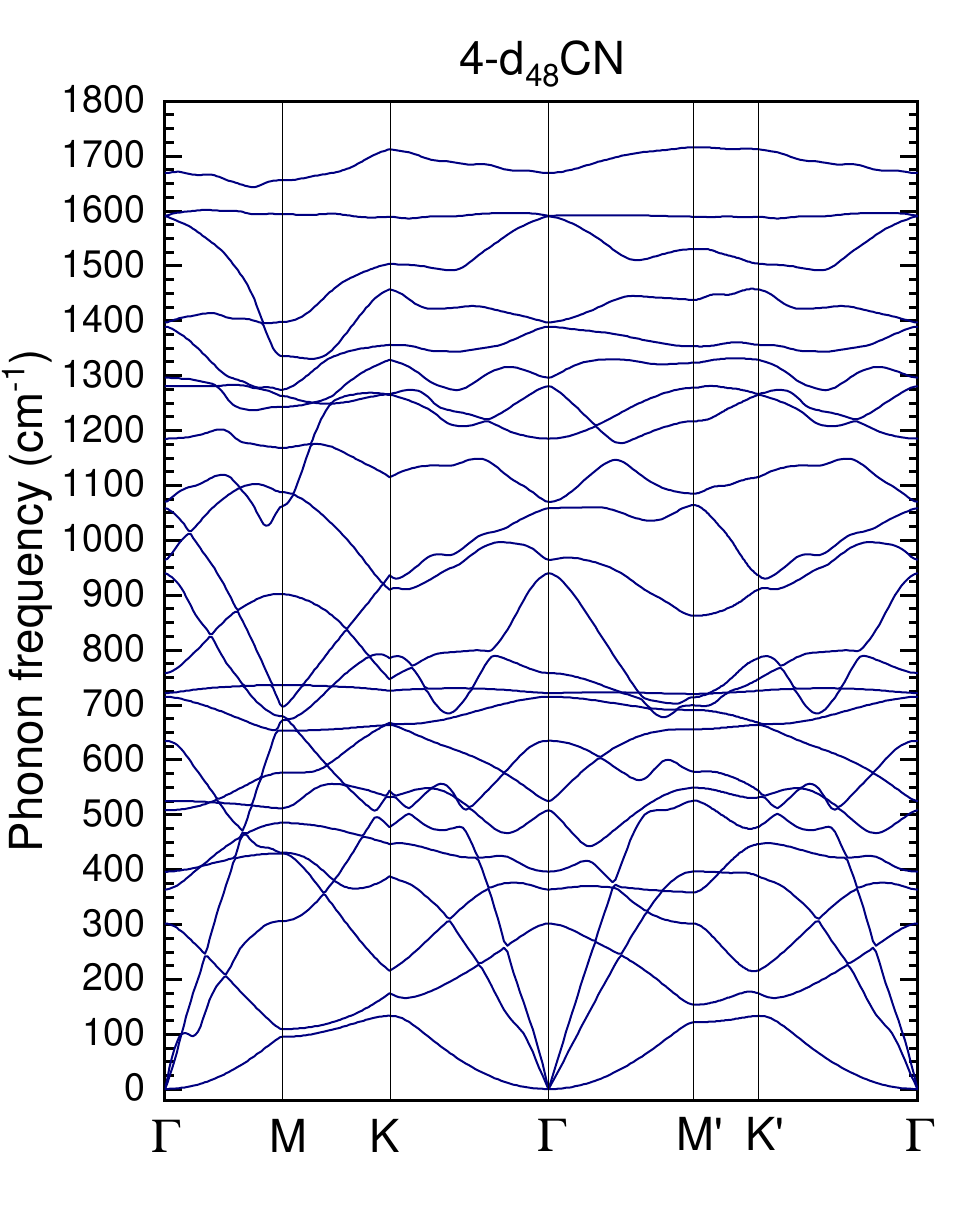}
\includegraphics*[width=0.32\columnwidth]{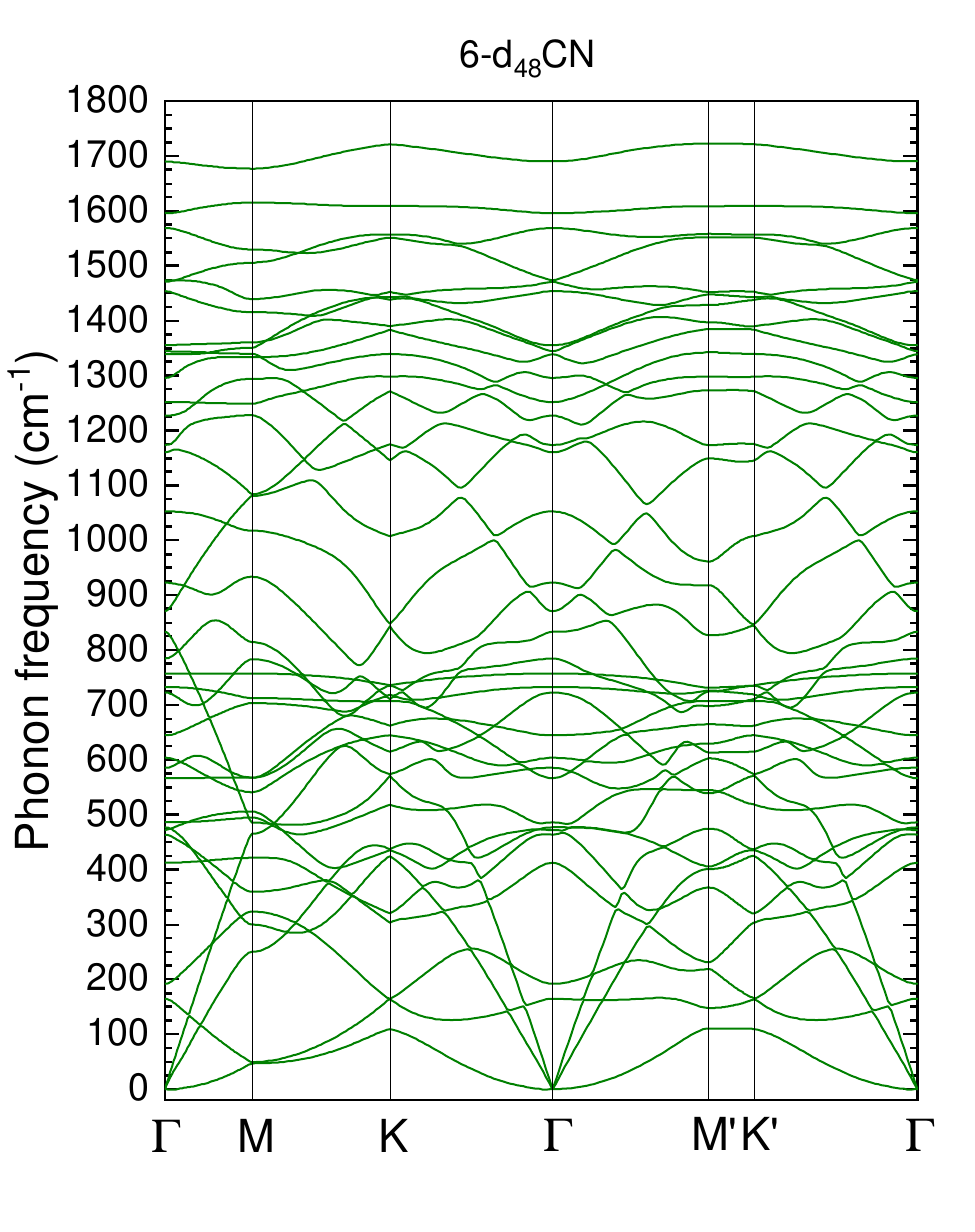}
\caption{\label{fig:phondisp} Phonon dispersions from the semiempirical GFN2-xTB method for selected defective networks.}
\end{figure}

Figure~\ref{fig:phondisp} shows the calculated phonon dispersions of 3-d$_{48}$CN, 4-d$_{48}$CN and 6-d$_{48}$CN networks, obtained using the GFN2-xTB method in combination with the ASE framework. For even (odd) $N$, we used 9x9 (9x12) supercells of the primitive cell to obtain the necessary force constants and then Fourier interpolated the dynamical matrix along the desired Brillouin zone path. For each of the force constants calculations, we used tight convergence thresholds of 10$^{-13}$\,eV and 10$^{-11}$ for the SCC error to reduce numerical noise and ensure accurate results. 
All phonon branches have positive frequencies, indicating that the network is dynamically stable despite the high defect line density. We expect that the network stability further increases as the defect line density is decreased (i.e. $N$ increases) and the nanoribbon components become larger and the networks become more graphene-like.

\FloatBarrier

\section{Interribbon bond lengths}
\begin{figure}[h!]
	\centering
	\includegraphics*[width=0.5\columnwidth]{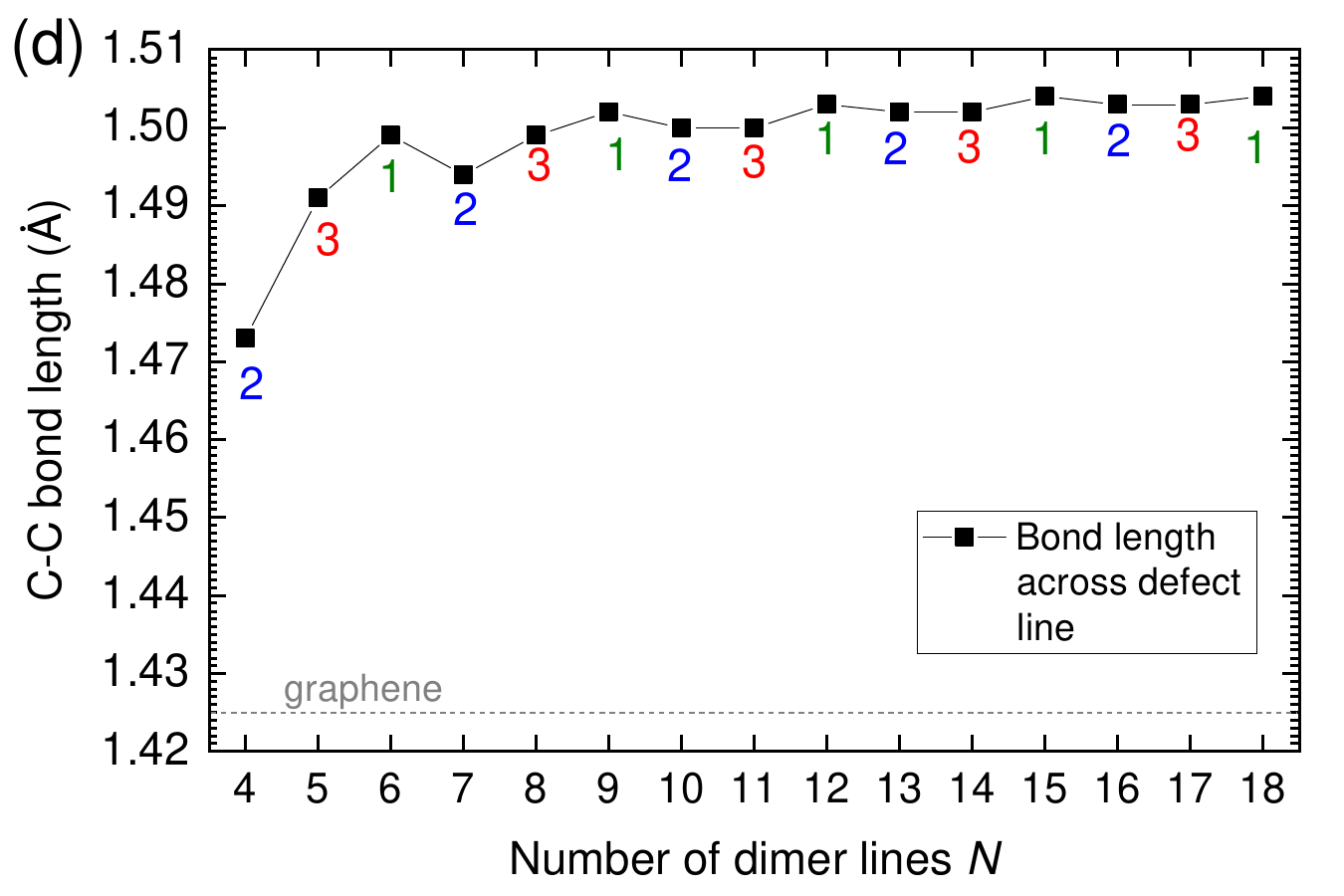}
	\caption{\label{fig:bondlengths} Bond lengths across the defect lines. Colored numbers indicate the network class.}
\end{figure}

Figure~\ref{fig:bondlengths} shows that the 'interribbon' bond lengths $d$ across the defect lines (thicker black lines in Fig. 1 of the main text) also exhibit a family behavior to a certain extent. For the highest defect density, the bond lengths are still rather short, 1.47\,\AA, due to the larger structural impact of the defect line, while the value quickly approaches 1.5\,\AA\space as $N$ increases.

\FloatBarrier
\newpage

\section{$N$-d$_{48}$CN bandstructures}
%

\begin{figure}[b!]
\centering
\includegraphics*[width=0.3\columnwidth]{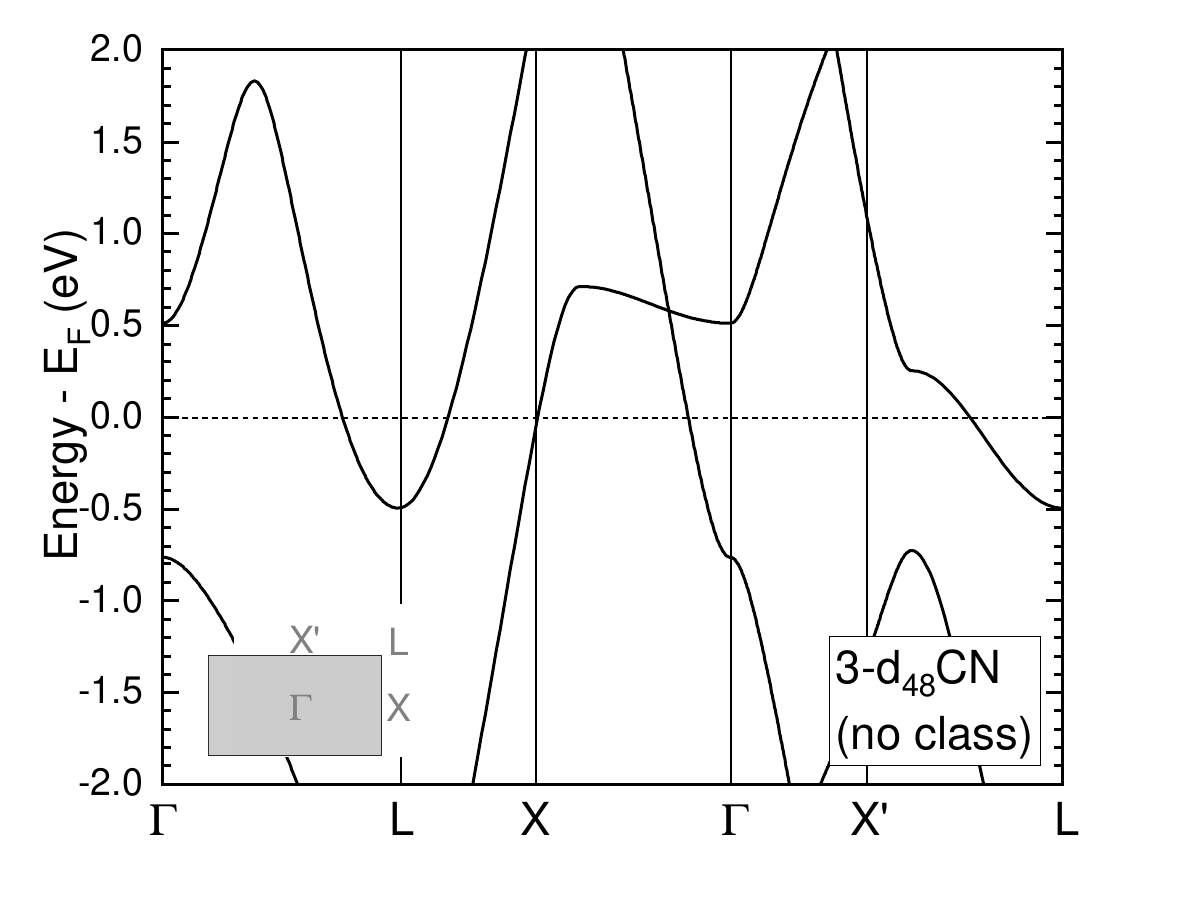}
\includegraphics*[width=0.3\columnwidth]{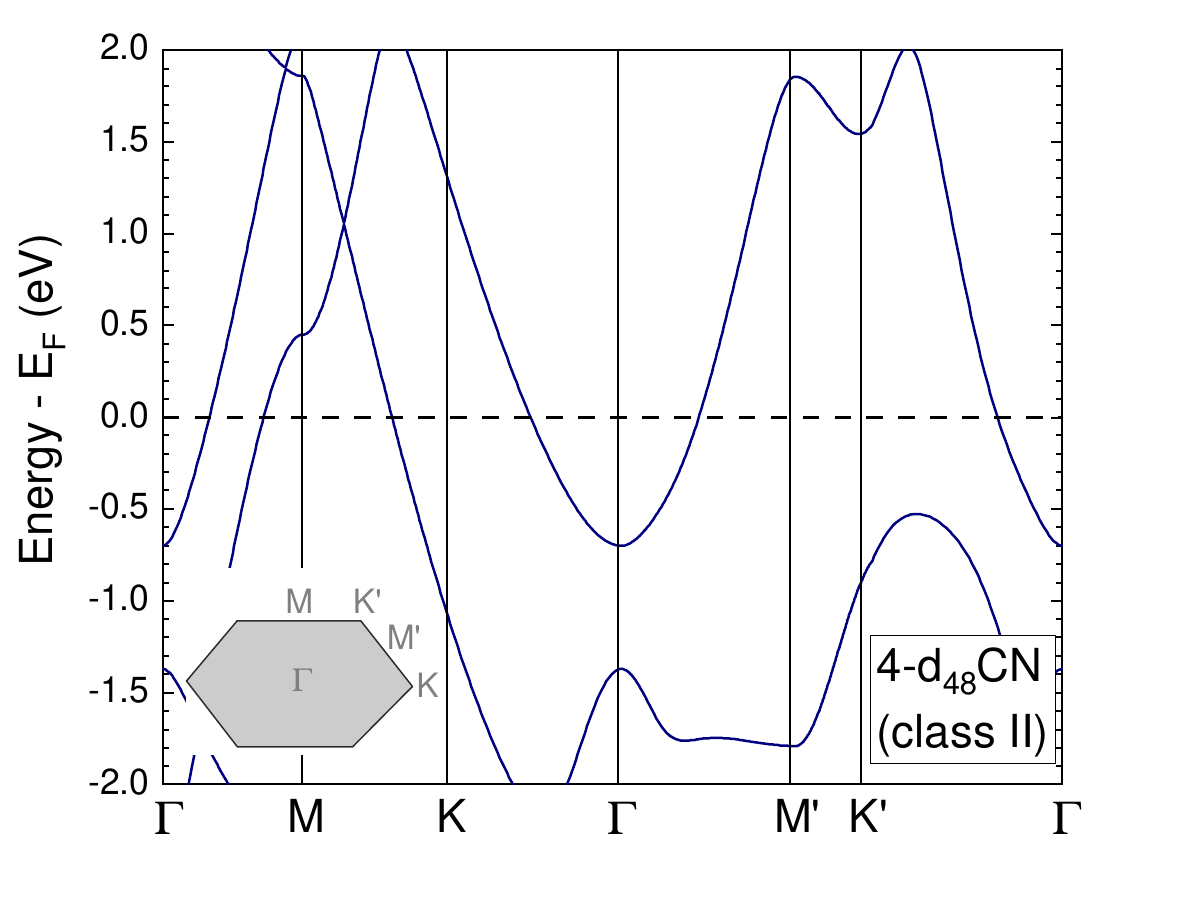}
\includegraphics*[width=0.3\columnwidth]{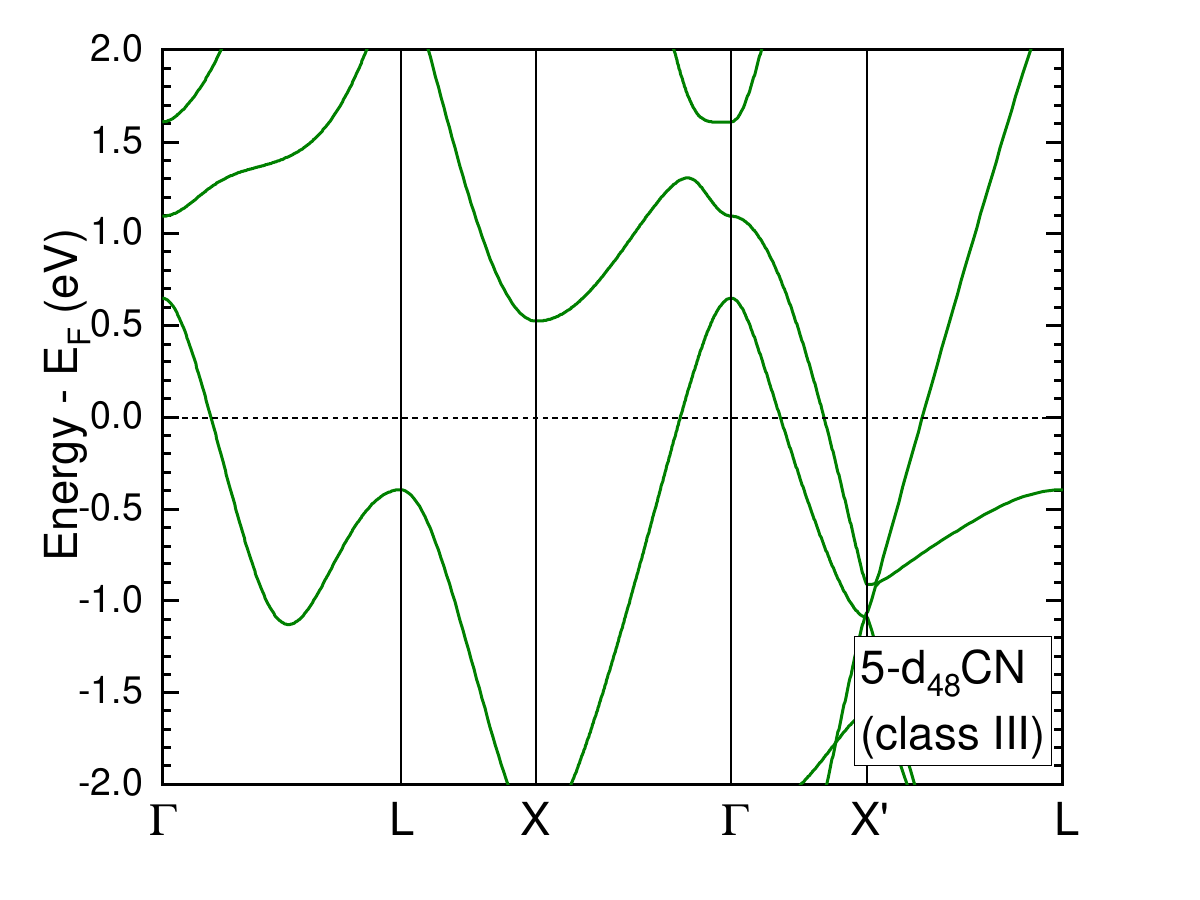}
\includegraphics*[width=0.3\columnwidth]{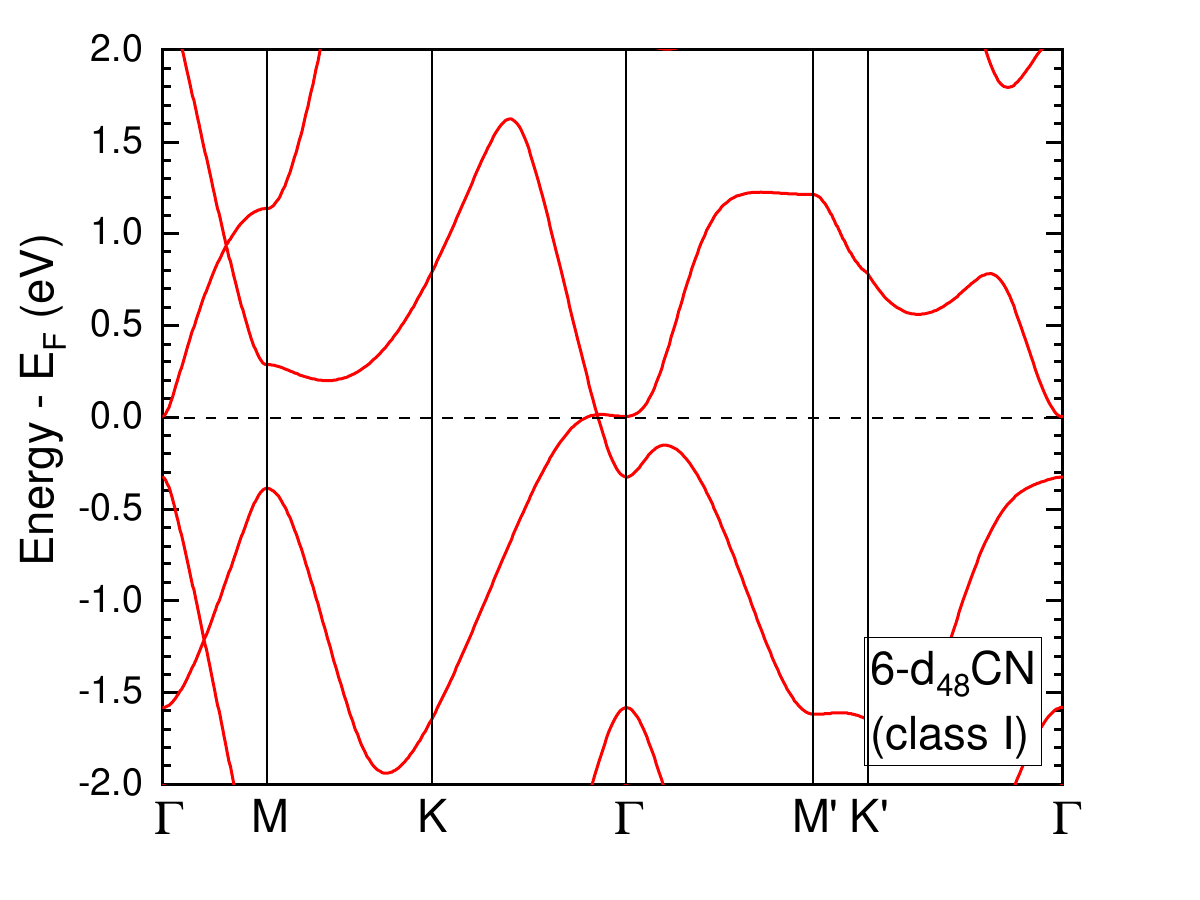}
\includegraphics*[width=0.3\columnwidth]{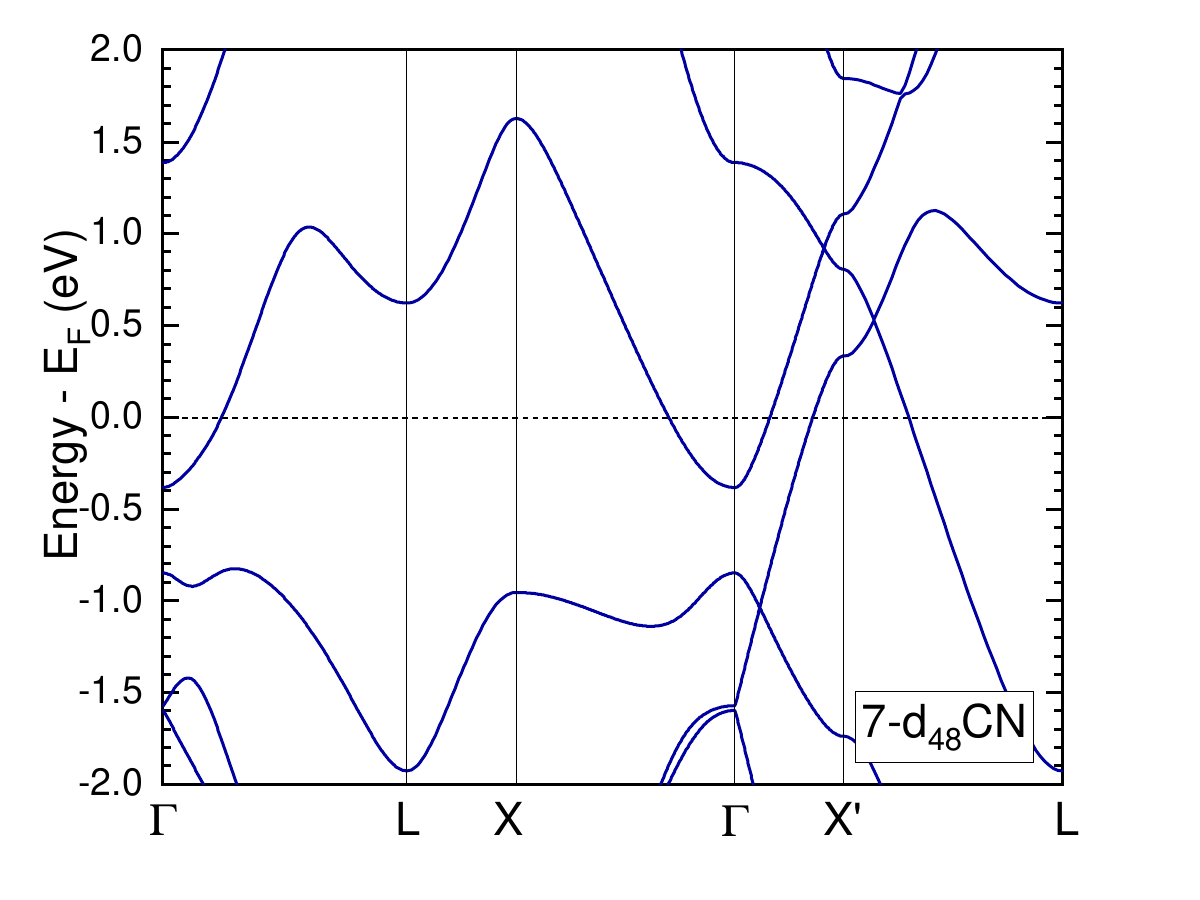}
\includegraphics*[width=0.3\columnwidth]{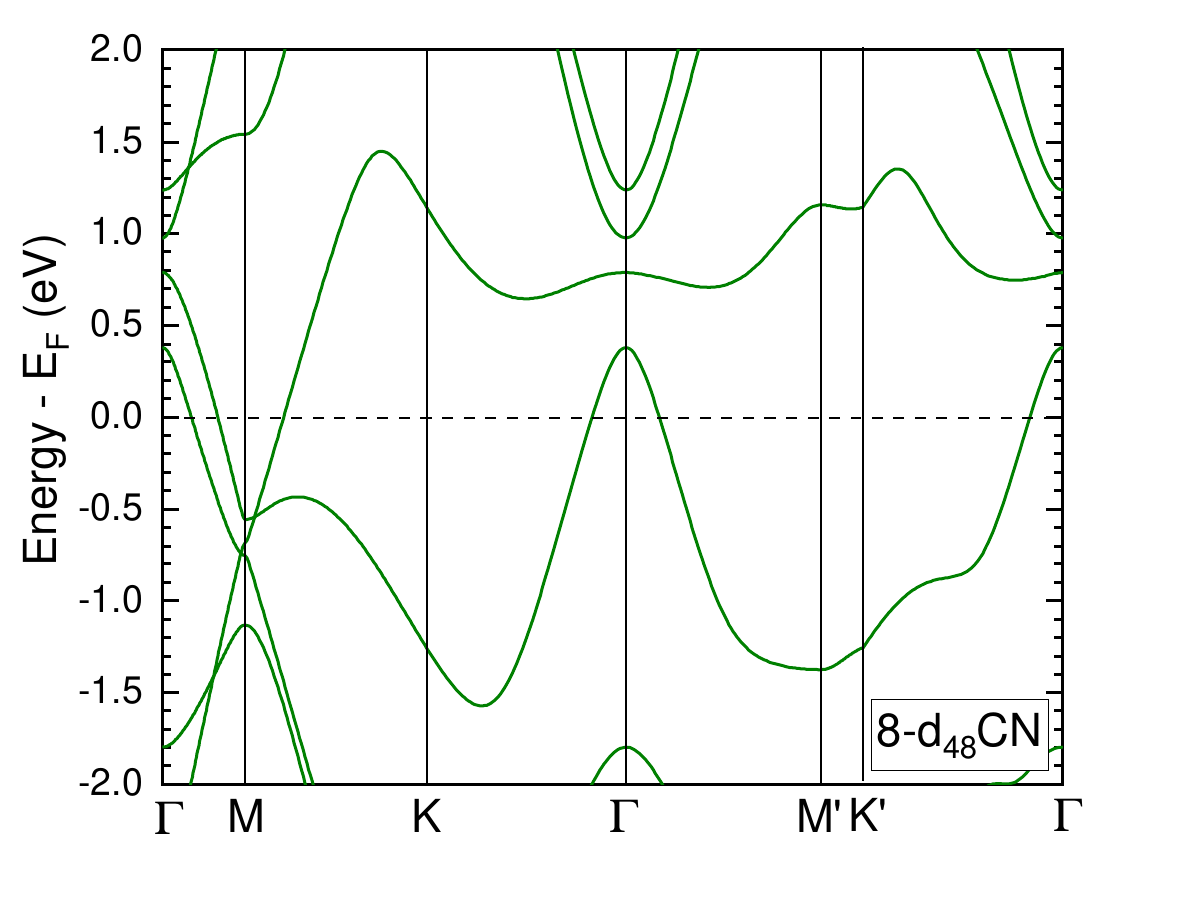}
\includegraphics*[width=0.3\columnwidth]{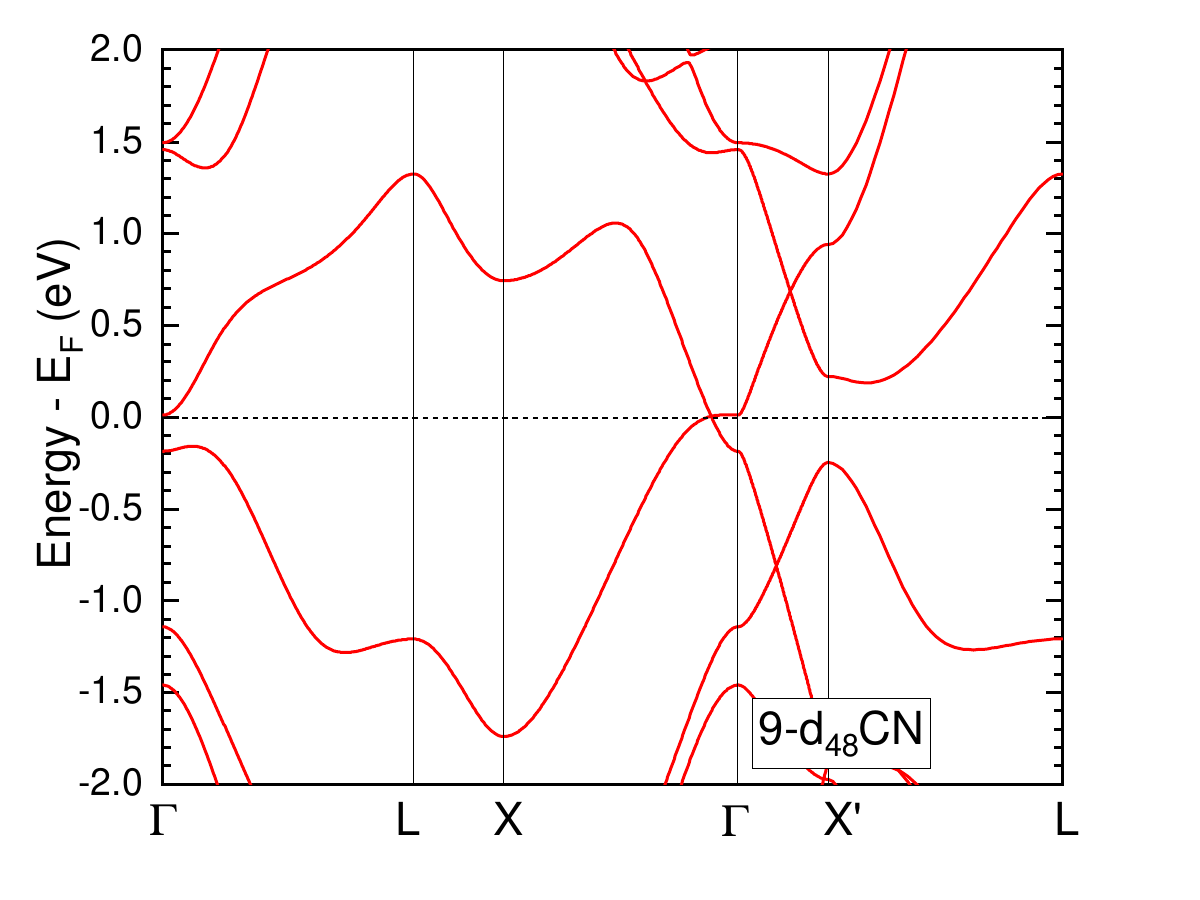}
\includegraphics*[width=0.3\columnwidth]{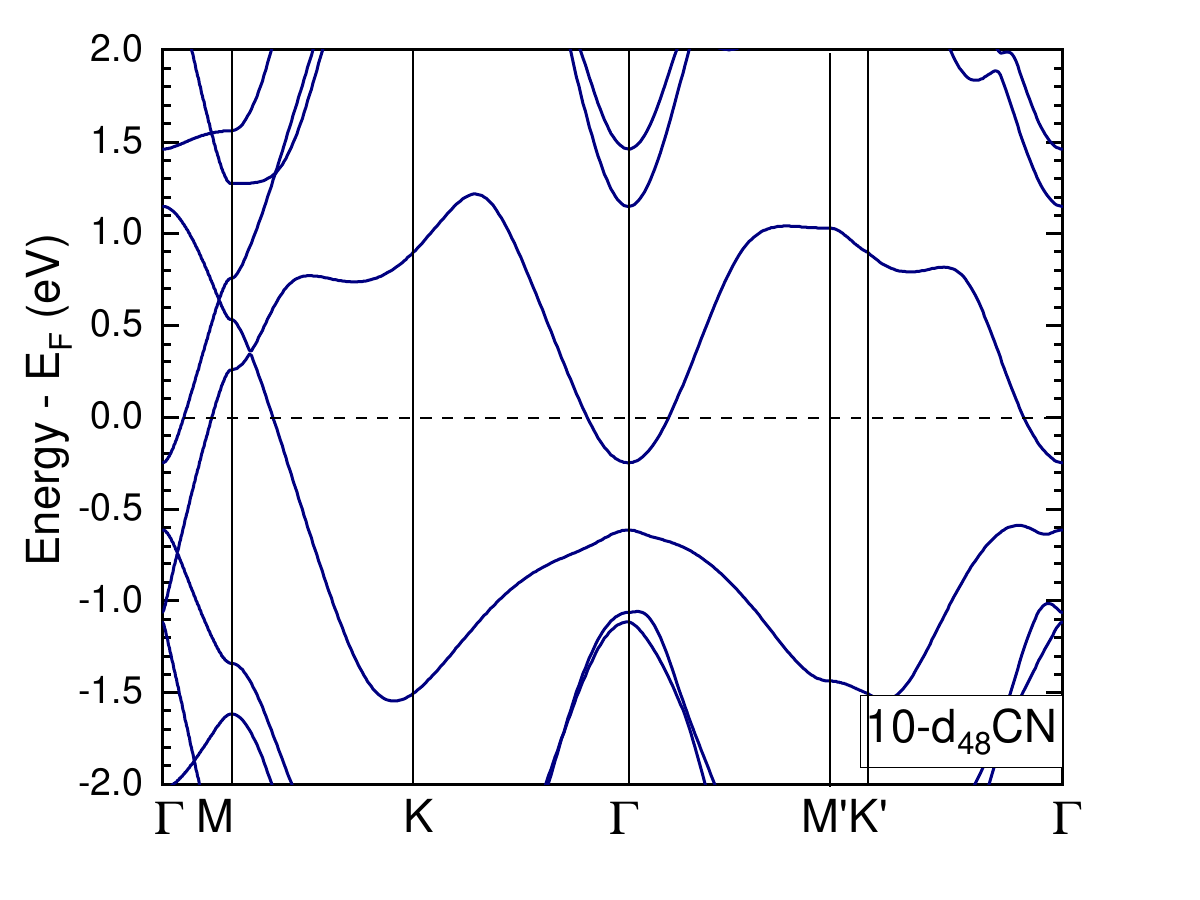}
\includegraphics*[width=0.3\columnwidth]{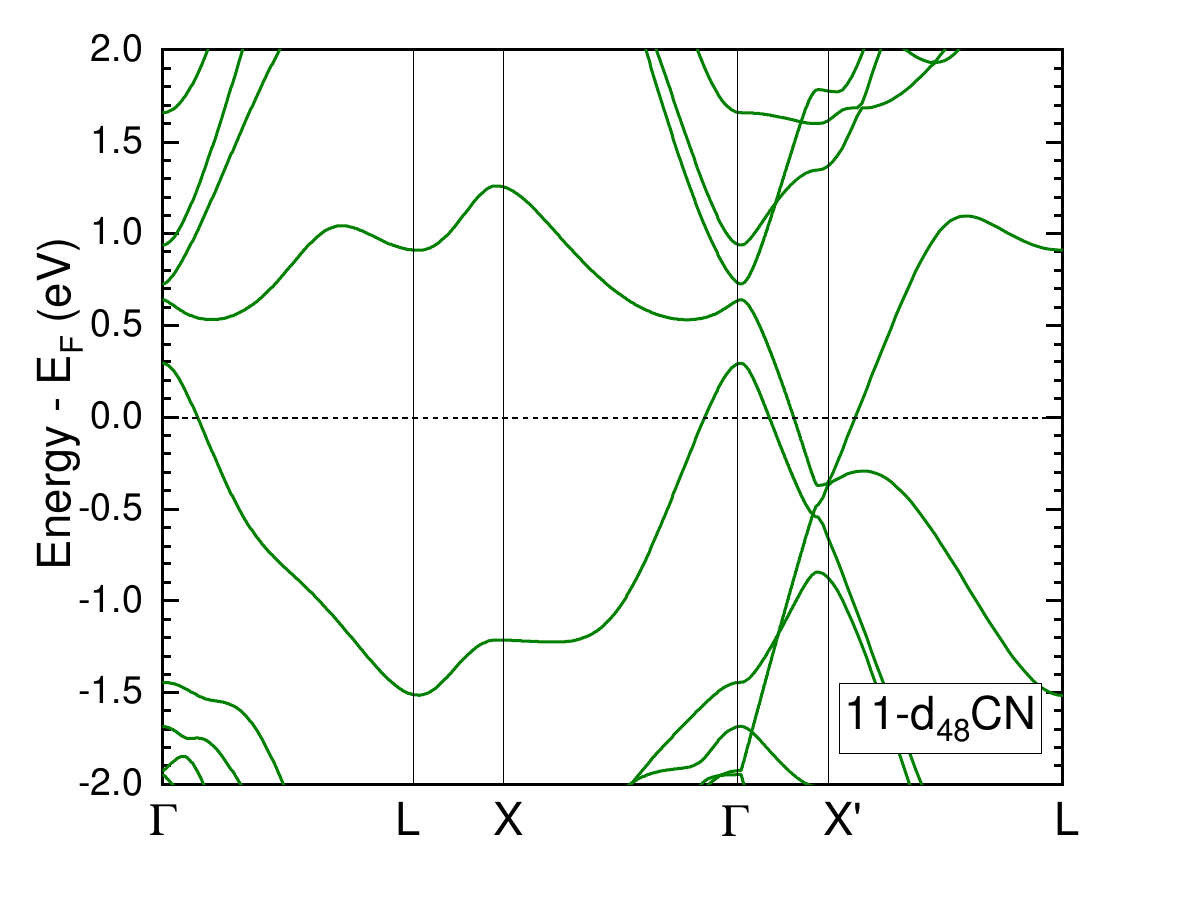}
\includegraphics*[width=0.3\columnwidth]{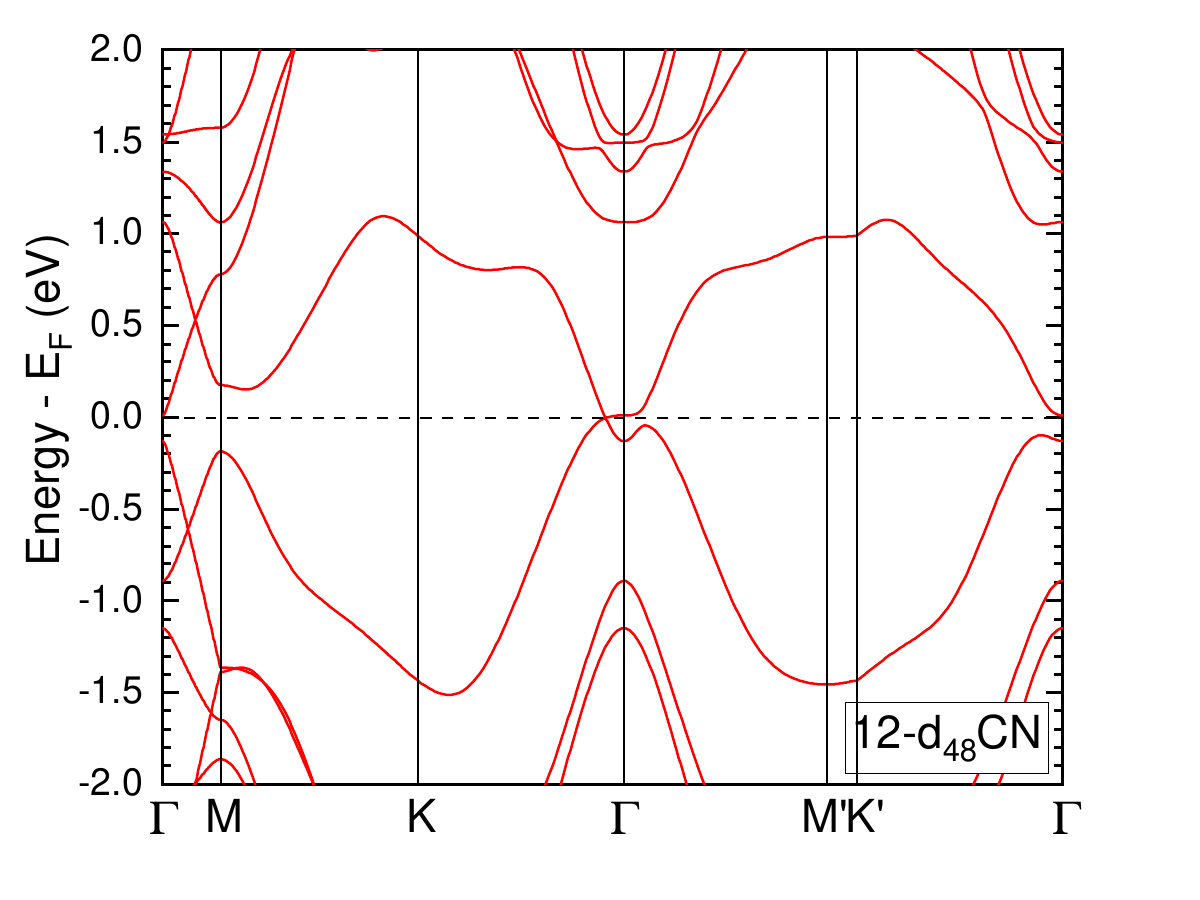}
\includegraphics*[width=0.3\columnwidth]{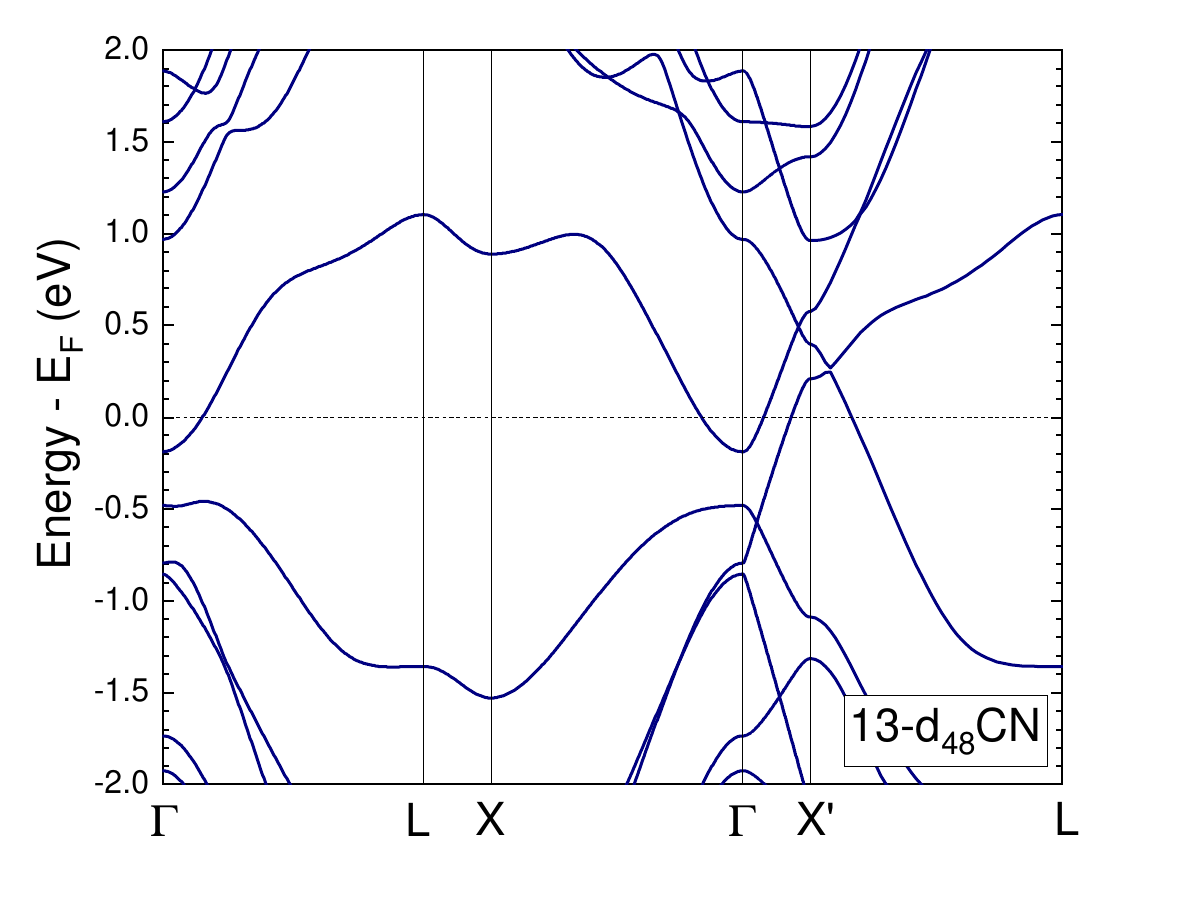}
\includegraphics*[width=0.3\columnwidth]{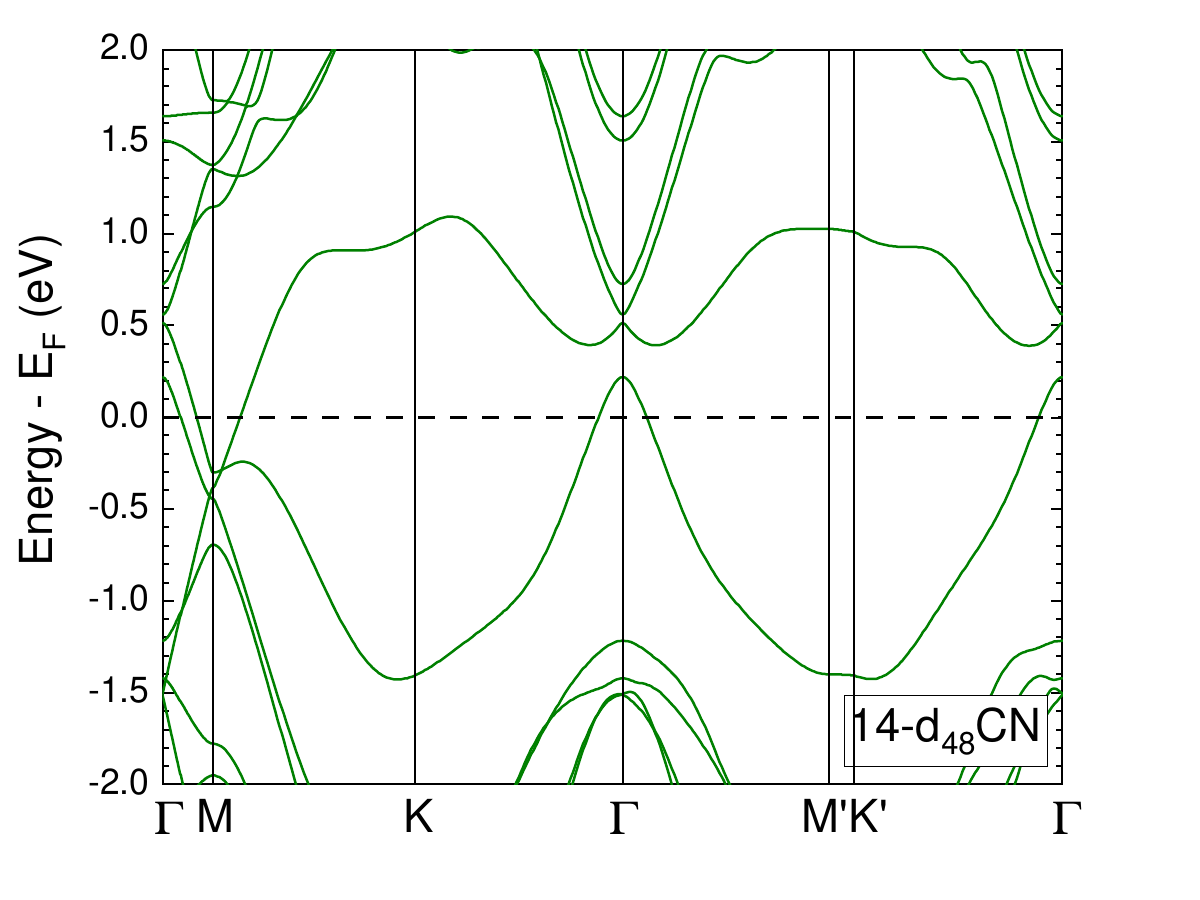}
\includegraphics*[width=0.3\columnwidth]{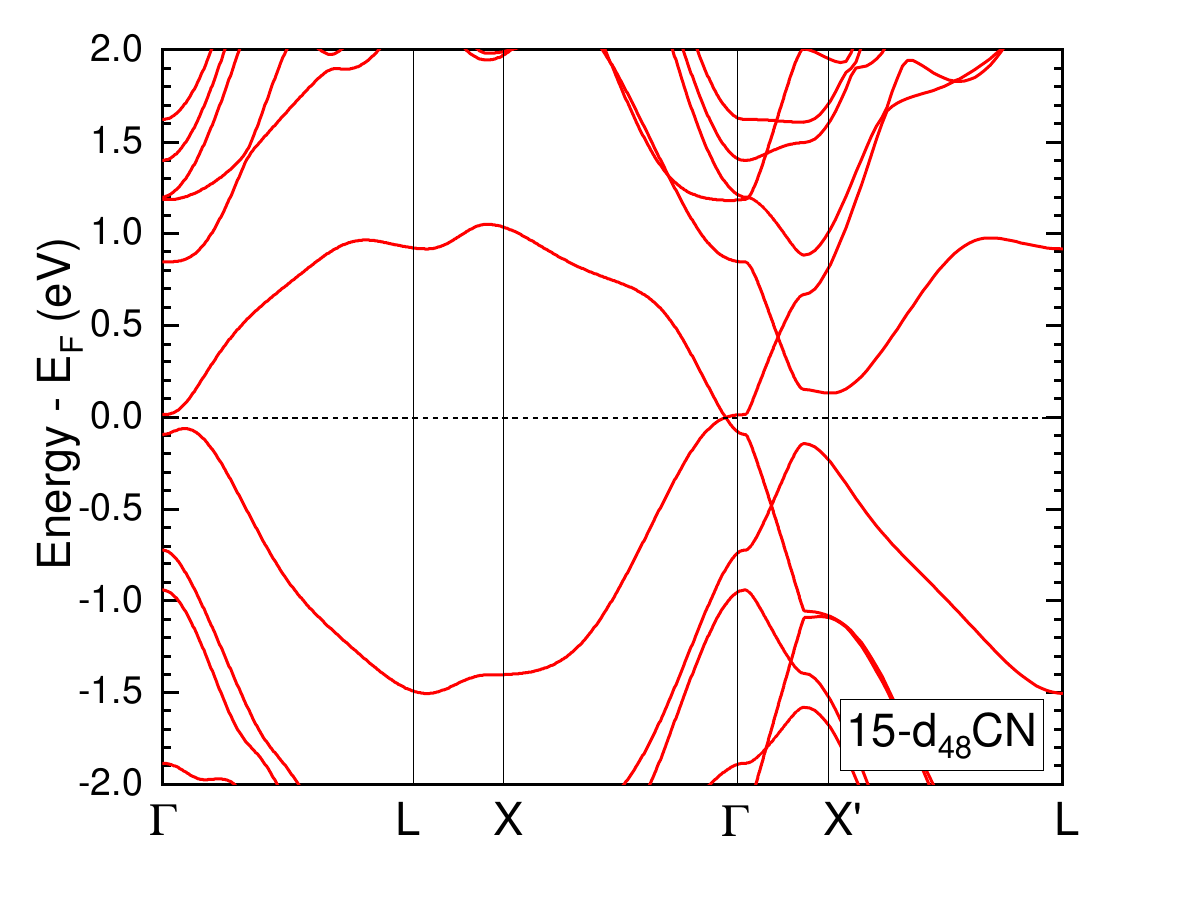}
\includegraphics*[width=0.3\columnwidth]{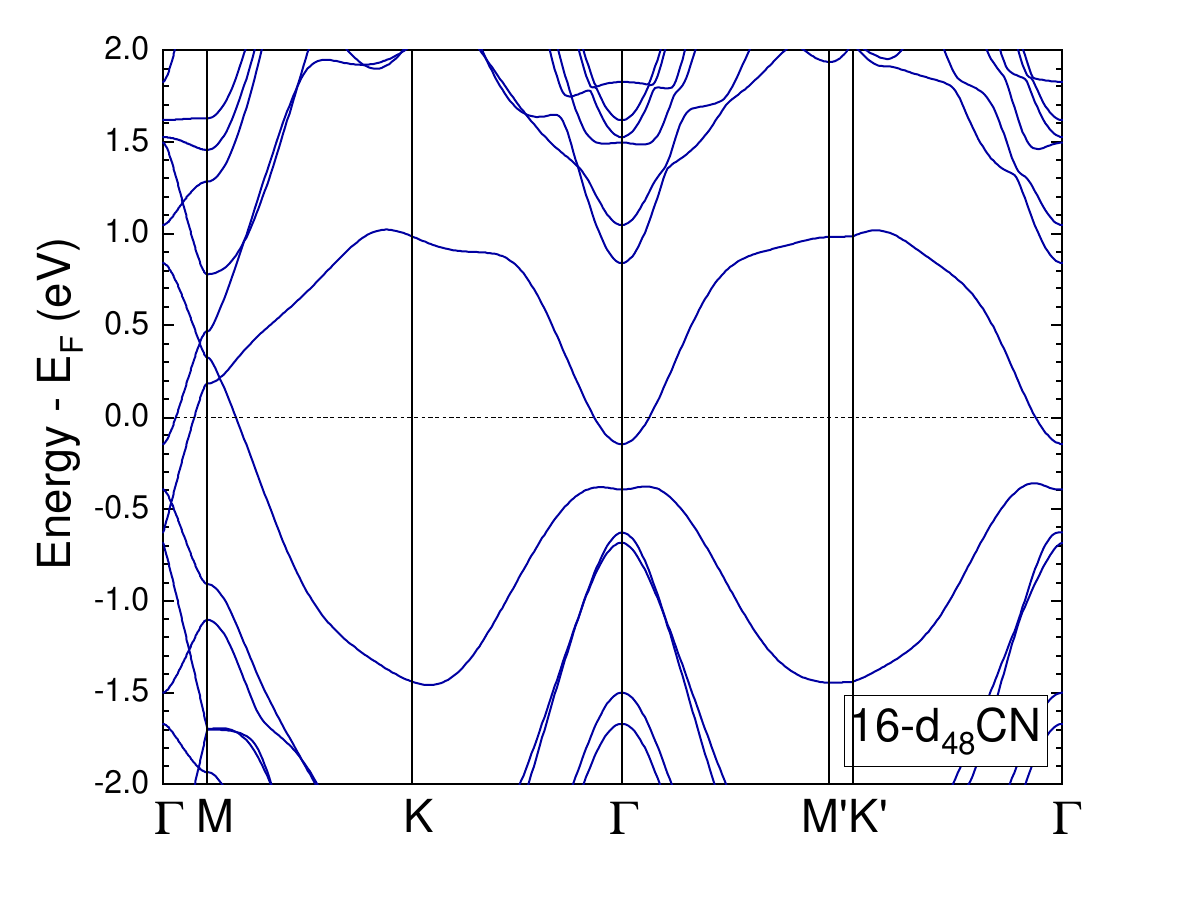}
\includegraphics*[width=0.3\columnwidth]{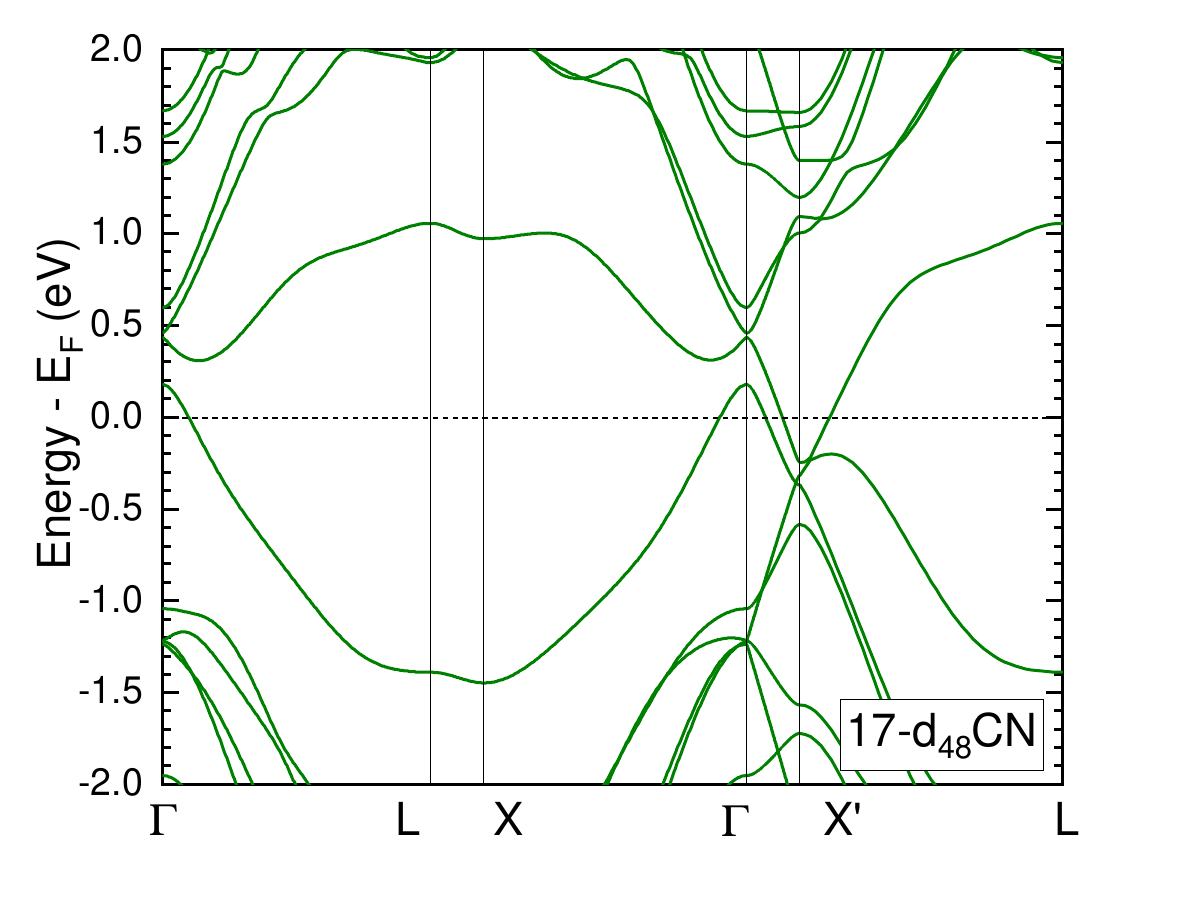}
\caption{\label{fig:bands_3-17} Electronic bandstructures of $N$-d$_{48}$CN for $N=3-17$ on the DFT-PBEsol level of theory. The left (red color), middle (blue color) and right (green color) columns show the bandstructures of class I, II and III networks, respectively. In all plots, the zero-of-energy was set to the Fermi energy of the system. }

\end{figure}

\FloatBarrier
\newpage

\section{Comparison of tight-binding bands}
\begin{figure}[h!]
\centering
\includegraphics*[width=0.45\columnwidth]{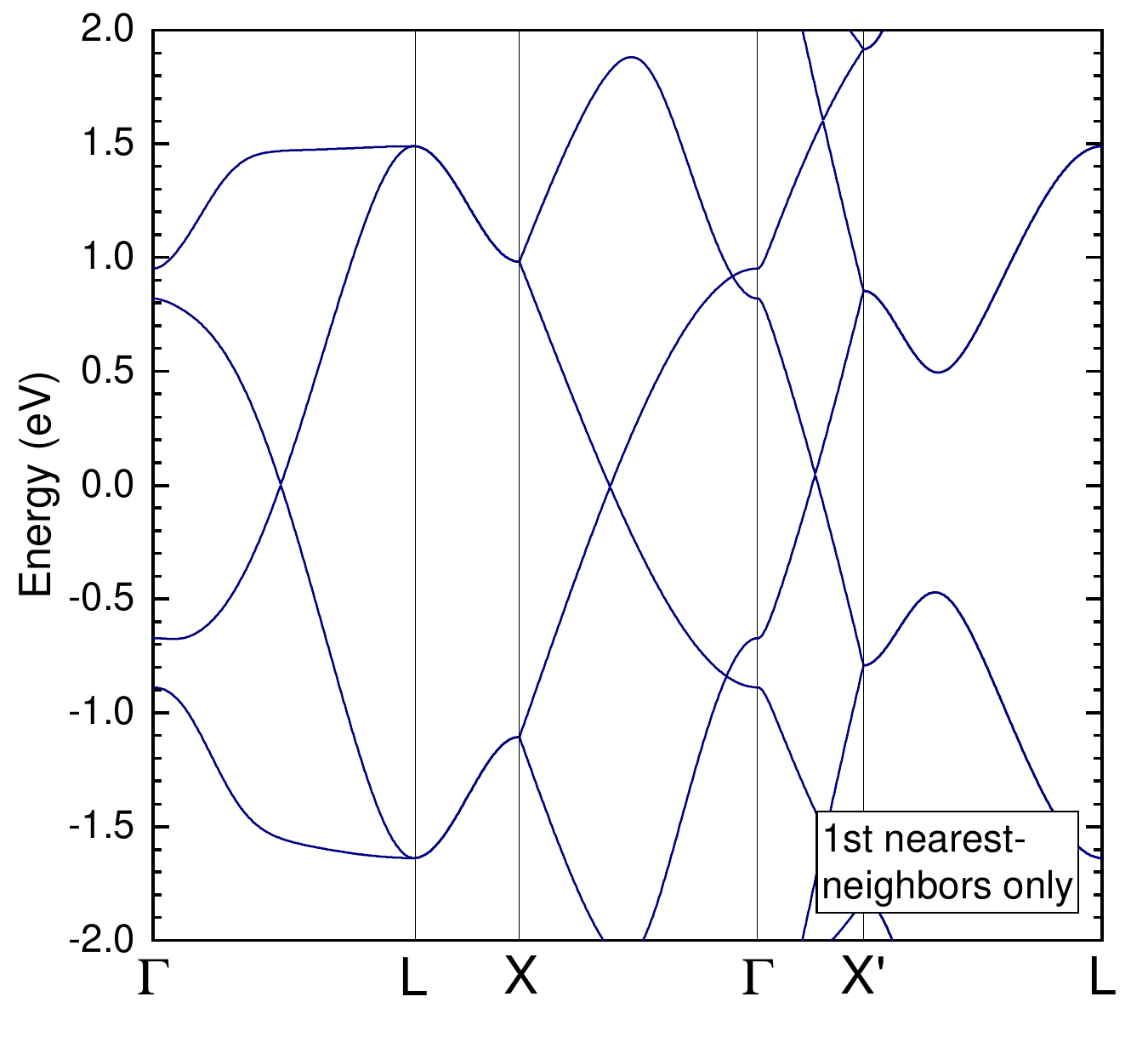}
\includegraphics*[width=0.45\columnwidth]{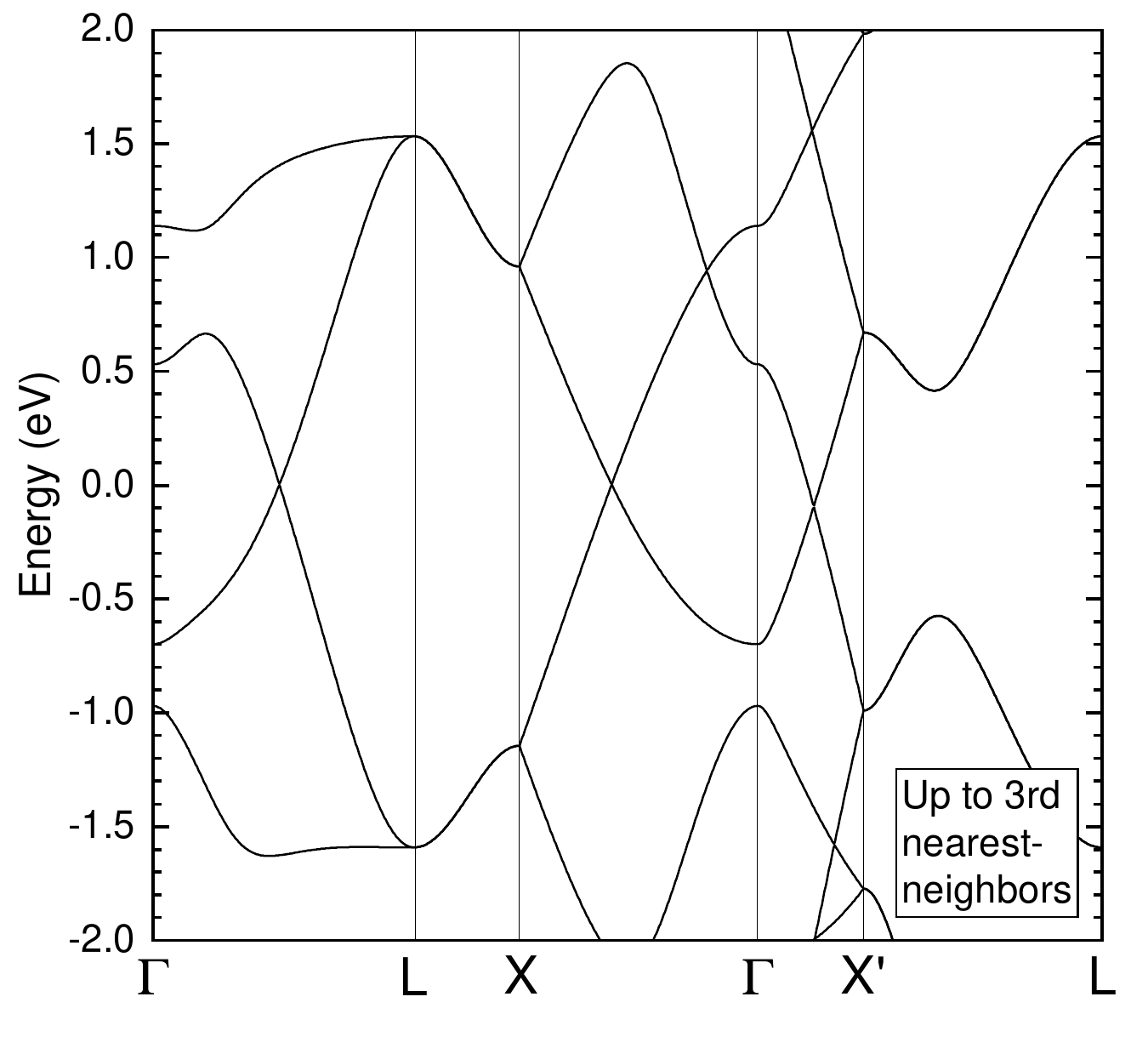}
\caption{\label{fig:TB} tight.binding bandstructures of a $4$-d$_{48}$CN for nearest neighbor (left side) and up to third-nearest neighbor coupling (right side) across the defect line. We used the fitted coupling parameters from defective nanoribbons [J. Phys.: Condens. Matter 36, 295501 (2024)], i.e. $\gamma_1=\gamma=2.68\,eV$ for nearest-neighbor coupling across the defect line, $\gamma_2=-0.35\,eV$ for second-nearest neighbor coupling, but a weaker $\gamma_3=0.1\,eV$ for third-nearest neighbor coupling. For the coupling within the nanoribbon segments, we used the values $t_1=-2.94\,eV$, $t_2=-0.3\,eV$, $t_3=-0.11\,eV$ from the same publication.}
\end{figure}

Figure~\ref{fig:TB} shows a comparison of the tight binding bandstructures of a 4-d$_{48}$CN for nearest- and up to third nearest neighbor hopping across the defect lines. Within the nanoribbon segments, up to third-nearest neighbor coupling is considered. 

Nearest-neighbor hopping across the defect line is sufficient to reproduce the linear band crossing at the Fermi energy and the general qualitative band dispersion. Inclusion of second- and third-nearest neighbor coupling across the defect line improves the reproduction of the band dispersion compared to DFT, at the cost of additional parameter fitting. 

\section{Convergence tests}
\begin{figure}[h!]
\centering
\includegraphics*[width=0.45\columnwidth]{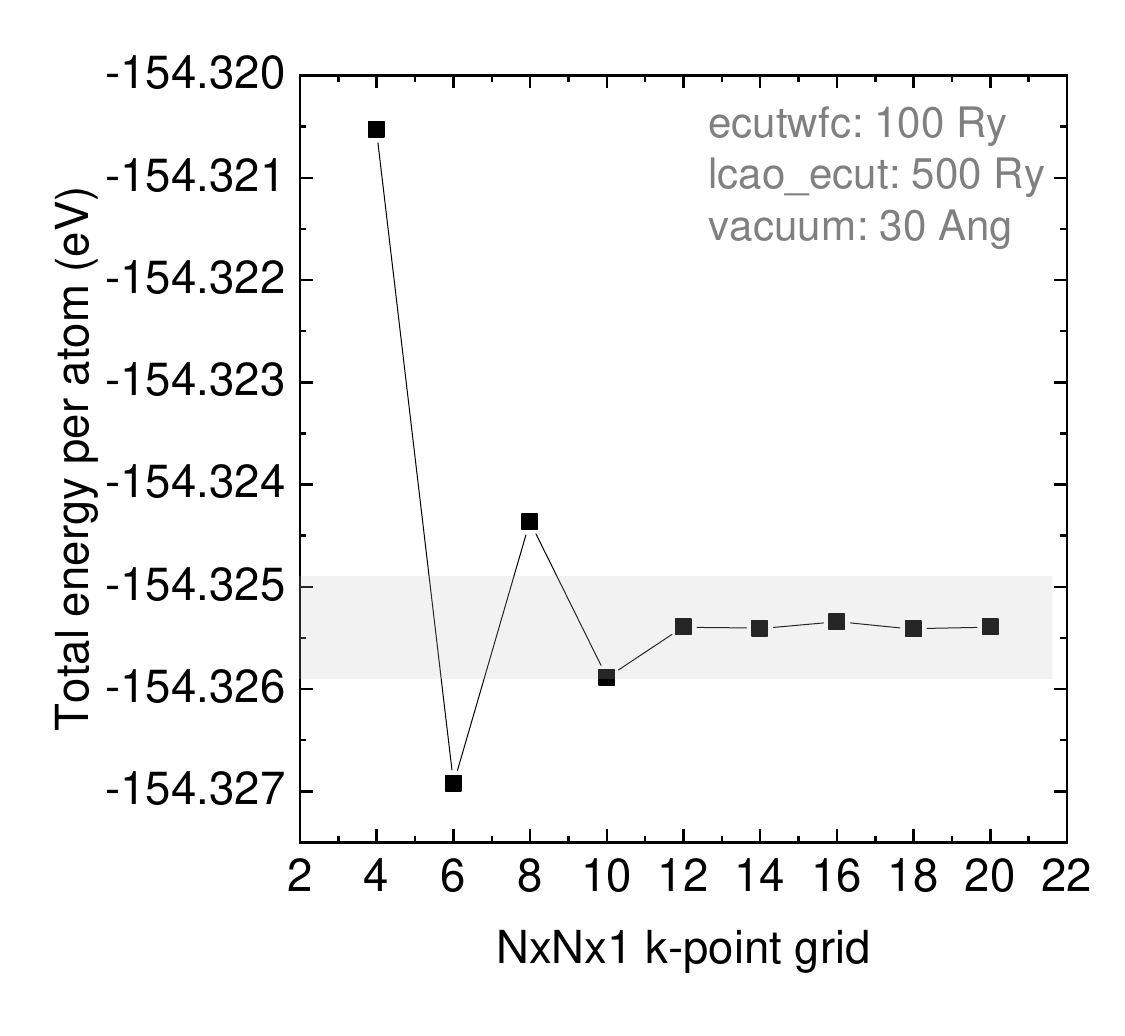}
\includegraphics*[width=0.45\columnwidth]{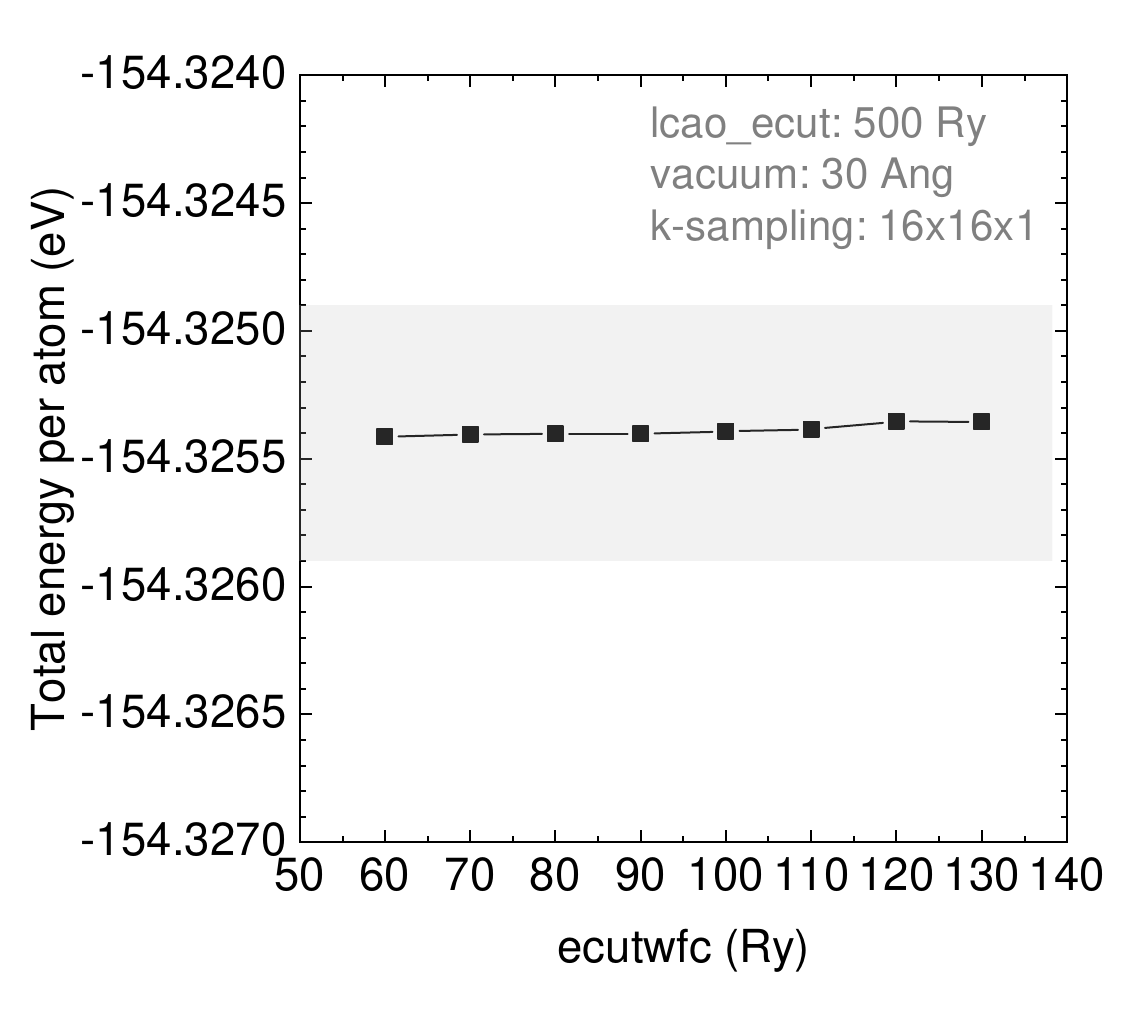}
\includegraphics*[width=0.45\columnwidth]{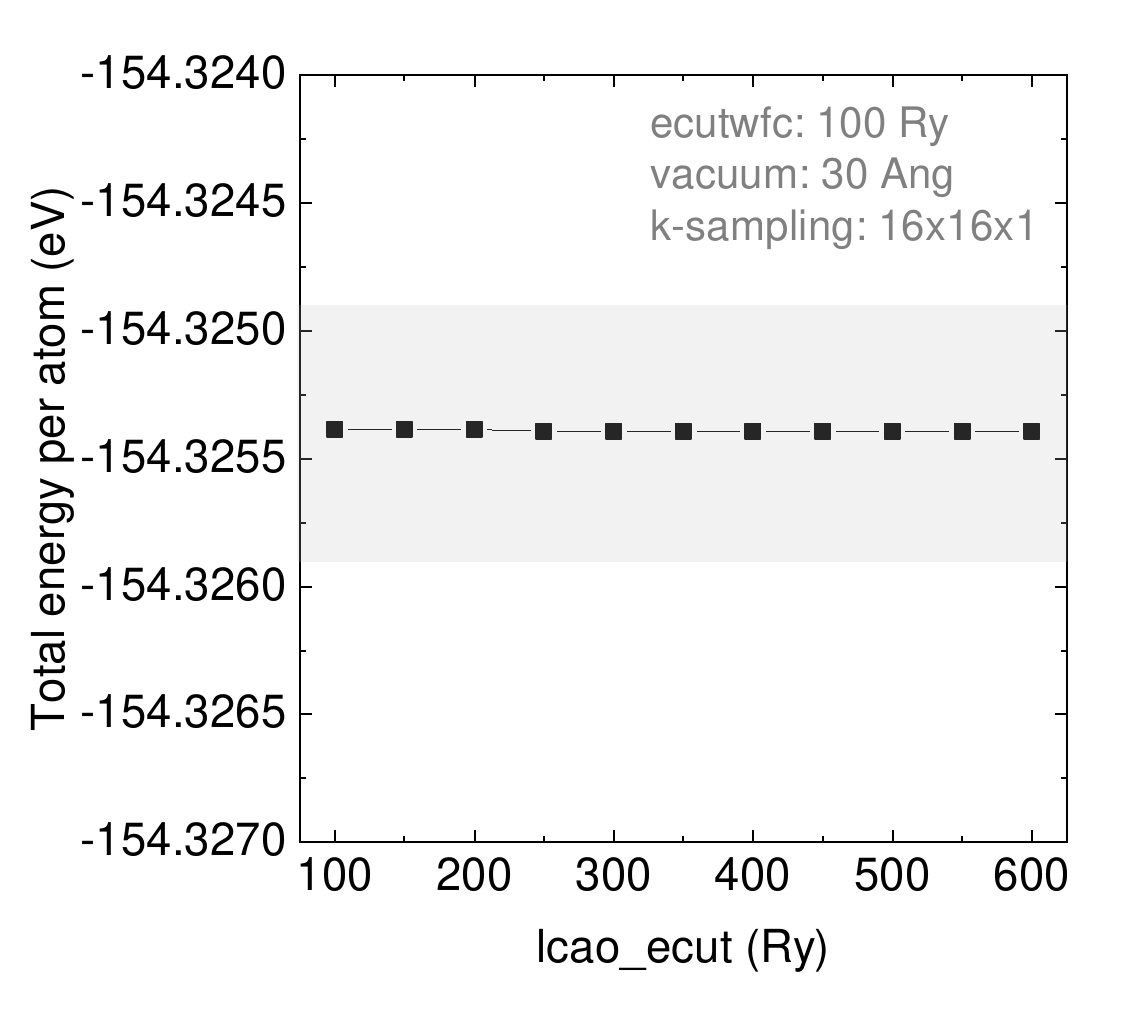}
\includegraphics*[width=0.45\columnwidth]{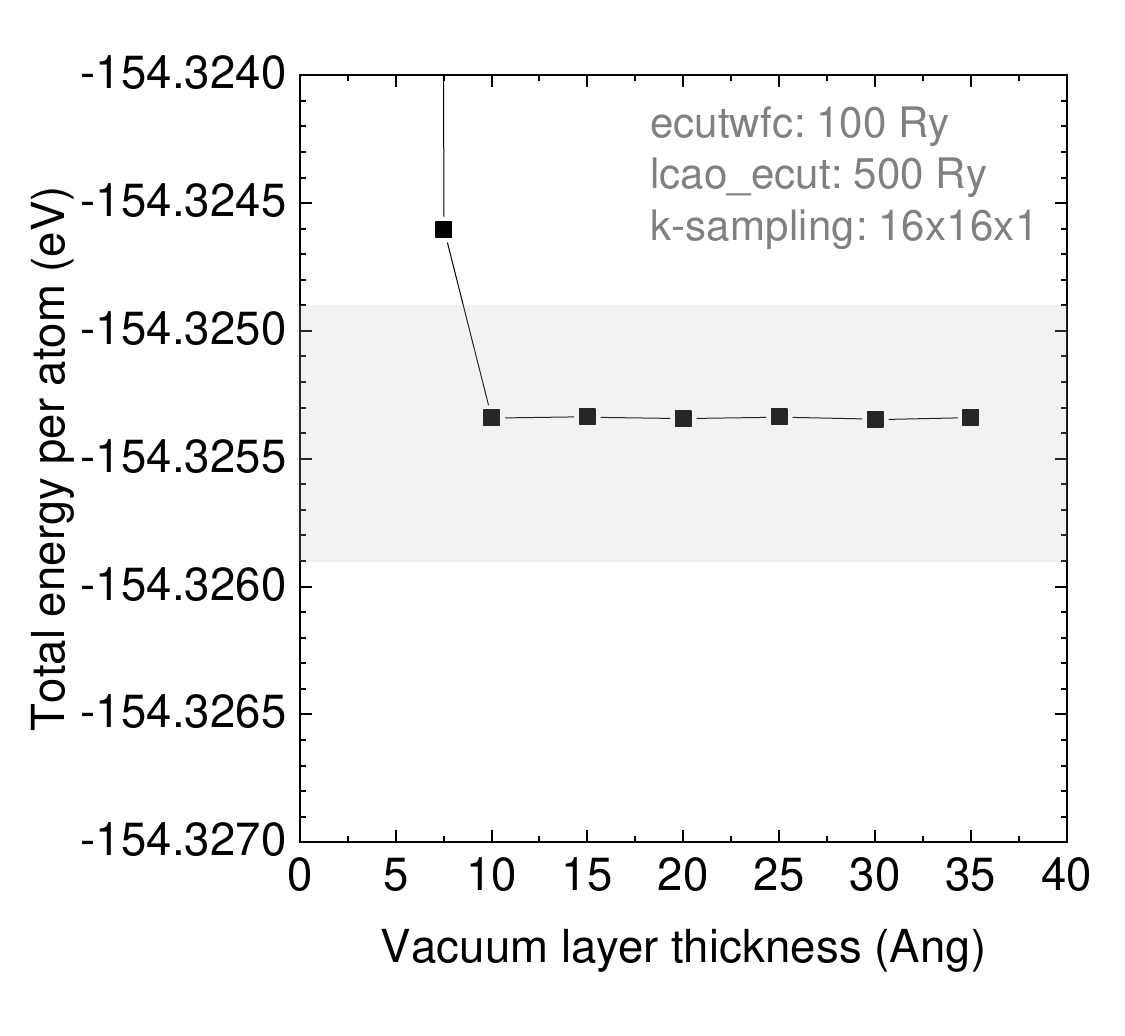}
\caption{\label{fig:convergence} Convergence tests for a 4-d$_{48}$CN.}
\end{figure}

Figure~\ref{fig:convergence} shows a the results from convergence tests for various  calculation parameters

\end{document}